\documentstyle[12pt,aaspp4]{article}

\begin{document}

\title{Prevalence and Properties of Dark Matter in Elliptical Galaxies}
\author{Michael Loewenstein\altaffilmark{1}}
\affil{Laboratory for
High Energy Astrophysics, NASA/GSFC, Code 662, Greenbelt, MD 20771}
\altaffiltext{1}{Also with the University of Maryland Department of Astronomy}
\author{Raymond E. White III}
\affil{Department of Physics \& Astronomy, University of Alabama,
Tuscaloosa, AL 35487-0324}

\begin{abstract}
Given the recently deduced
relationship between X-ray temperatures and stellar velocity
dispersions (the ``$T$--$\sigma$ relation'')
in an optically complete sample of elliptical galaxies
(\cite{dw96}), we demonstrate that $L>L_*$
ellipticals contain substantial amounts of dark matter {\it in general}.
We present constraints on the dark matter scale length and
on the dark-to-luminous mass ratio within the optical
half-light radius and within the entire galaxy. 
For example, we find that minimum values of dark matter core radii
scale as $r_{\rm dm} > 4(L_V/3L_*)^{3/4}h^{-1}_{80}$ kpc and that
the minimum dark matter mass fraction is $\gtrsim$20\% within one optical
effective radius $r_e$ and is $\gtrsim39$--$85$\% within $6r_e$, depending on
the stellar density profile and observed value of $\beta_{\rm spec}$.
We also confirm the prediction of Davis \& White (1996) that the dark
matter is characterized by velocity dispersions that are greater
than those of the luminous stars: $\sigma_{\rm dm}^2\approx1.4$--$2\sigma_*^2$.
The $T$--$\sigma$ relation
implies a nearly constant mass-to-light ratio within 
six half-light radii: $M/L_V\approx 25h_{80}$M$_{\odot}/$L$_{V_\odot}$.
This conflicts with the simplest extension of CDM theories of large
scale structure formation to galactic scales; we consider a couple
of modifications which can better account for the observed 
$T$--$\sigma$ relation.

\end{abstract}


                                \clearpage
\section{Background}

There is a strong consensus that spiral galaxies are embedded
in massive, non-luminous halos, the structure
and scaling properties of which have
been intensively investigated (e.g., \cite{pss}).
Studies of galaxy kinematics and of the density and temperature
distributions of hot intergalactic gas
in galaxy groups (\cite{m96}) and clusters (e.g., \cite{m98})
have demonstrated the dominance of dark matter on larger scales.
In the context of existing theories of the formation
of galaxies and large scale structure, it is therefore natural to expect
non-luminous material within elliptical galaxies as well.
However, until very recently, the evidence for dark matter
in ellipticals has been less forthcoming and more controversial.

Dark matter in ellipticals has been elusive largely due to the lack of
internal dynamical indicators at optical and radio wavelengths
that are as ubiquitous, spatially extensive, and unambiguous
as HI rotation curves in spirals. Although mergers have endowed
some early-type galaxies (many classified as S0) with
neutral (e.g., \cite{fvd}) or ionized (\cite{b93}) gas disks, 
these are rare and
may not be representative of the population of ellipticals as a whole.
Studies of stellar kinematics have, until recently, been inadequate
in quality and extent to significantly constrain the
large scale mass distribution in ellipticals in light of the complicating
effects of projection and anisotropic velocity distributions.

Most, if not all, giant elliptical galaxies contain
extended distributions of hot X-ray emitting gas
that provide a means for probing the large
scale dark matter distribution --- a means potentially
as powerful as rotation curves in spiral galaxies.
Since these hot gaseous halos
should be close to hydrostatic equilibrium (\cite{lm87}), measurements
of the density and temperature distributions from X-ray
imaging and spectroscopy
yield constraints on the mass distribution on the scale and with
the accuracy and spatial resolution of X-ray
temperature profiles. Observations of 
elliptical galaxies by the {\it Einstein Observatory}
IPC were presented as strong evidence for the presence
of extensive dark halos in early-type galaxies (\cite{fjt}; \cite{f86};
\cite{lm87}); however, the temperature profiles were of sufficiently
poor quality to foster a continuing skepticism
(\cite{f89}, \cite{df91}).

Progress on a number of fronts has been rapid during recent
years. In the optical, improvements in detector capabilities, 
data analysis, and modeling
have now provided robust stellar dynamical constraints on the
mass distribution out beyond the optical half-light radius $r_e$
for several ellipticals
(\cite{s93}, \cite{c95}, \cite{r97}, \cite{g98}), while gravitational lensing
observations are providing the first evidence for dark matter
in intermediate redshift galaxies (\cite{k95}; \cite{bbs}; \cite{g96}).
X-ray observations using the {\it ROSAT} and {\it ASCA} satellites
have improved the accuracy, spatial resolution, and extent of derived
hot gas density and temperature profiles. Of particular interest is the recent
study of {\it ASCA} observations of $\approx 30$ early-type galaxies by
Matsushita (1997) that shows that gas temperature profiles are relatively
flat out as far as can be reliably measured (5--$20^\prime$, depending on
the intrinsic luminosity and galaxy distance). The existence of isothermal
0.5--1 keV gaseous halos extending, in some cases, to well beyond 10$r_e$
is strong evidence for the presence of extended dark matter in
elliptical galaxies.

The case for dark matter halos in a number of elliptical galaxies
is now overwhelming, and the following pattern seems to be emerging
from the latest results on individual galaxies: dark matter comprises a
significant but not dominant fraction of the total mass
within $r_e$ (\cite{s93}; \cite{dis}; \cite{r97}; \cite{g98}; but,
see \cite{m94} for an exceptional case: NGC 4636); 
dark matter becomes increasingly
dominant at large radii: $M/r\approx 5\times10^{12}$M$_{\odot}$ kpc$^{-1}$
to within a factor of two for $r>30$ kpc
(\cite{m94}, \cite{is96}, \cite{bc97}), corresponding to mass-to-light
ratios $>100$M$_{\odot}/$L$_{B_\odot}$
measured at (or extrapolated to) $\sim100$ kpc (\cite{bld}; \cite{g96}).

\section{Motivation and Overview}

While there is now little doubt that some
elliptical galaxies do contain
dark matter, a number of broader issues remain to be addressed. Among the
most important with respect to our understanding of galaxy formation is
whether dark matter in ellipticals is ubiquitous, and
how the relative scaling of dark and luminous matter varies --- both within
galaxies as a function of radius, and between galaxies with
different optical properties.
In this paper we use the recently
derived collection of X-ray temperatures in a complete, optically selected
sample from Davis \& White (1996; hereafter DW) to constrain the
{\it average} properties of dark halos in the {\it population} of
$L>L_*$
elliptical galaxies.

DW analyzed X-ray spectra from 42 of the 43 optically
brightest elliptical galaxies, using
{\it ROSAT} PSPC data if available and {\it Einstein} IPC data otherwise
({\it i.e.} for four galaxies). See DW and White \& Davis (1998) for details
about the sample and data analysis. For the purposes of this work,
we use only the mean relationship between X-ray temperatures,
$\langle T\rangle$,
and central projected optical velocity dispersions, $\langle\sigma\rangle$
(hereafter, the ``$T$--$\sigma$ relation'') ---
$\langle T\rangle\propto {\langle\sigma\rangle}^{1.45}$ --- or equivalent
X-ray/optical correlations.
More precisely, we
consider the variation of the observable $\beta_{\rm spec}$
with optical luminosity
$L_V$ (Figure 1), where
\begin{equation}
\beta_{\rm spec}\equiv
{{\mu m_p\langle\sigma\rangle}^2\over {k\langle T\rangle}},
\end{equation}
and $\mu m_p$ is the mean mass per particle.
If dark matter dominates
the gravitational potential on large scales, then
$\langle T\rangle$ is essentially a measure of the dark matter content
within the extraction radii used by DW ($6r_e$, where $r_e$ is the optical
effective radius).
Since the observed scaling relations ({\it i.e.} the
fundamental plane) for elliptical galaxies provide a link between
$\langle\sigma\rangle$ and the global luminosity,
$\beta_{\rm spec}$ is an excellent diagnostic of
the dark-to-luminous matter ratio within the optical radius.

Temperature profiles can provide additional dark matter constraints,
but are measurable for only the X-ray brightest galaxies in the sample.
Likewise, extended stellar velocity dispersion profiles are useful, but 
data of sufficient detail to be unambiguously interpreted are not
generally available. On the other hand,
central velocity dispersions have been accurately measured for the
entire sample; central velocity dispersions are known to correlate with 
other optical properties and are less likely to suffer from
the uncertainties caused by anisotropic stellar orbits.

Our goal in this paper is to make general inferences about the presence of
dark matter halos and their mean systematic variation with optical luminosity.
Toward that end, we utilize a set of physically and observationally
well-motivated mass models and scaling relations in order to reproduce the 
$T$--$\sigma$ relation. Although we address the 
scatter in the $T$--$\sigma$ relation,
a detailed examination of individual galaxies is
beyond the scope of this paper and is deferred to future investigation.

Our results
can be considered in three stages.
First, we demonstrate
how the observed range of $\beta_{\rm spec}$
{\it necessarily} requires the existence of dark halos in elliptical galaxies.
Second, we derive conservative lower limits
on the dark matter scale
length and dark-to-luminous mass fraction, and discuss how the dark matter
content must scale with optical luminosity
so as to reproduce the observed $T$--$\sigma$ relation.
Finally, we discuss possible implications of our
results for cosmology and galaxy formation, by embedding our
models within particular scenarios for large scale structure formation.

\section{Modeling}

As stated above, we use the observable $\beta_{\rm spec}$,
defined in equation (1), to constrain the
dark matter distribution in elliptical galaxies. This parameter
is the ratio of suitable averages (to be defined below)
of the square of the stellar velocity dispersion and the gas
temperature. Therefore, we
must derive the expected distributions of stellar velocity dispersions
$\sigma$ and gas temperatures $T$, given specified total gravitational
mass distributions.

\subsection{The Stellar Component}

Recently published {\it Hubble Space Telescope} ({\it HST})
observations of the
centers of elliptical galaxies show that their surface brightness profiles
can generally be characterized by the function
\begin{equation}
I(R)=2^{(b-c)/a}I(r_{\rm br})\left({{r_{\rm br}}\over R}\right)^c
\left[1+\left({R\over {r_{\rm br}}}\right)^a\right]^{(c-b)/a},
\end{equation}
where $R$ is the projected radius and
$r_{\rm br}$ is a break radius such that
$I\propto R^{-c}$ for $R\ll r_{\rm br}$ and $I\propto R^{-b}$ for $R\gg r_{\rm br}$,
while $a$ characterizes the sharpness of the break
(\cite{f97}).
In order to accurately calculate the central stellar velocity dispersion,
this observed departure from distributions with ``analytic cores''
must be accounted for.
In the Appendix we demonstrate that, for $a=2$ and $c<1$, equation (2)
provides a good approximation to the projection of a space
density profile $\propto r^{-d}$ ($c<d<c+1$) for
$r\ll r_{\rm br}$ and $\propto r^{-b-1}$ for $r\gg r_{\rm br}$.
To derive the luminosity density
distribution $l(r)$, we numerically deproject the surface brightness profile:
\begin{equation}
l(r)={{2^{(b-c)/a}I(r_{\rm br})}\over {\pi r_{\rm br}}}
{\int_0}^{x^{-1}}dz(1-x^2z^2)^{-1/2}z^b(1+z^a)^{(c-b-a)/a}
(cz^a+b),
\end{equation}
where $x\equiv r/r_{\rm br}$, and the change of variables
$z\equiv r_{\rm br}/R$ has been made.

The DW sample is magnitude-limited, and we are
primarily interested in the most luminous ellipticals, since
they have the most accurate X-ray temperatures (see \S 3.3).
We adopt the typical values for $M_V<-22$
galaxies (adopting $H_0=80$ km s$^{-1}$ Mpc$^{-1}$)
of $a=2$, $b=1.44$, and $c=1/10$ (\cite{f97}).
For the nine galaxies in common with DW (excluding the highly 
flattened NGC 4697), 
the mean surface brightness parameters and variances
from Faber et al. (1997) are as follows:
$a=1.88\pm 0.51$, $b=1.38\pm 0.16$, $c=0.096\pm 0.076$.
The resulting
stellar density profile diverges slightly faster than $r^{-1}$ for
$r\ll r_{\rm br}$.
The asymptotic surface brightness slope $b$ is flatter than in 
conventional surface brightness parameterizations and the
integrated luminosity $L(r)$ does not converge. Therefore, we
truncate the optically luminous distribution at radius
$r_{\rm max}$, determined so that
\begin{equation}
L_{\rm core}\equiv \pi r_{\rm br}^2I(r_{\rm br})=0.012L(r_{\rm max});
\end{equation}
consequently
$r_{\rm max}=200r_{\rm br}$.

As an alternative stellar model, we consider the density profile
from Hernquist (1990),
\begin{equation}
l(r)={L\over {2\pi}}{{r_{\rm Hern}}\over r}{1\over {(r+r_{\rm Hern})^3}},
\end{equation}
where $L$ is the total luminosity and $r_{\rm Hern}=0.45r_e$ (as adopted here)
provides the best match to a de Vaucouleurs (deV) model with effective
radius $r_e$
(\cite{h90}). An advantage of this model is that the density and mass
profiles are analytic. We shall 
subsequently refer to the stellar distribution
of equation (3) as the ``HST'' model, and that of equation (5)
as the ``Hernquist'' model.
A comparison of the luminosity density (in units of
the average density, $3L/4\pi{r_{\rm max}}^3$) as a function of normalized
radius ($r/r_{\rm max}$) in these models with
the corresponding deV model is shown in Figure 2, where
$r_{\rm max}=200r_{\rm br}=40r_{\rm Hern}/3=6r_e$. All three models produce
similar slopes at very small $r$; however, deV and Hernquist
models fail to reproduce the flattening at $\approx r_{\rm br}$ observed by
{\it HST}. On the other hand, while $\approx 80$\%
of the total luminosity is contained within $\approx 0.7r_{\rm max}\approx 4r_e$
in  all three models, the 
HST model concentrates the entire
remaining $\approx 20$\% of the mass
between $\approx 4r_e$ and $6r_e$ as opposed to spreading it out
from $\approx 4r_e$ to $\infty$.
In order to accurately characterize the observed central stellar
properties as required by the definition of $\beta_{\rm spec}$
without introducing additional parameters
({\it i.e.} an additional break radius and a distinct asymptotic form
for the density profile), such
a redistribution of the stellar light is necessary.
If the light profile
does indeed steepen outside of the {\it HST} field-of-view, then
the only effect on our results for the HST model is that,
by overestimating the stellar mass at large radii,
we would underestimate the dark halo mass required to reproduce any
observed value of $\beta_{\rm spec}$.

The stellar velocity dispersion is derived by solving the Jeans equation
in the form
\begin{equation}
{d\over {dr}}\left(\rho_*{\sigma_r}^2\right)=-{{2A}\over r}\rho_*{\sigma_r}^2
-{{GM(<r)}\over {r^2}}\rho_*,
\end{equation}
where $\rho_*$ is the stellar density distribution obtained from multiplying
the luminosity density (see equations 3 and 5, above) by a constant stellar
mass-to-light
ratio; $M(<r)$ is the total mass within radius $r$
(the dark matter component of which is described in the next section);
$A$ is the velocity dispersion anisotropy parameter
($A\equiv 1-{\sigma_t}^2/{\sigma_r}^2$); and $\sigma_r$ ($\sigma_t$) is the
radial (tangential) component of the velocity dispersion. We make the
reasonable assumption that the velocity dispersion becomes isotropic
at small $r$ but radial at large $r$
(\cite{cp92}), and adopt the functional form
\begin{equation}
A={{r^2}\over {r^2+s^2}},
\end{equation}
where the transitional scale radius $s$ is a free parameter
of our models.

\subsection{The Dark Matter Component}

The dark matter mass distribution
is assumed to follow the ``universal'' density profile
(\cite{nfw}),
\begin{equation}
\rho_{\rm dm}\propto\left({r\over {r_{\rm dm}}}\right)^{-1}
\left(1+{r\over {r_{\rm dm}}}\right)^{-2},
\end{equation}
that, when integrated, yields
\begin{equation}
M_{\rm dm}\propto f(r/r_{\rm dm})\equiv
\left[ {\rm ln} {{r+r_{\rm dm}}\over {r_{\rm dm}}}-{r\over {r+r_{\rm dm}}}\right].
\end{equation}

It is not clear how appropriate equations (8) and (9) are for elliptical
galaxy dark halos. On one hand, there is evidence for cores in spiral galaxy
dark matter distributions (e.g., \cite{pss}); on the other, dissipational
collapse of the baryonic component in ellipticals can increase the dark
matter concentration. Moreover, models with cores cannot generally
be consistently coupled with stellar mass distributions as cuspy as those
often observed in elliptical galaxies (\cite{cp92}). 
We have verified the existence of self-consistent 
two-component mass models
with dark matter following equation (8) and Hernquist stellar models
(equation 7),
in the parameter range of interest, according to the 
criterion of Ciotti and Pelligrini (1992).

From the virial theorem it follows that the integrated mass can
be robustly estimated from
the mean temperature.
Constraints on
the dark matter concentration can also be placed from
consideration of the 
$T$--$\sigma$ relation; however, any inferences
about dark matter at small or large radii are contingent on the corresponding
assumed asymptotic dark matter density slopes.
Our choice of
equation (8) for the dark matter
distribution does allow us to connect our results with predictions of
dark halo structure from numerical structure formation
simulations. However, the true
shape of the dark matter spatial distribution
can only be constrained from temperature and stellar velocity dispersion
{\it profiles} of sufficient extent and resolution.

\subsection{The Hot Gas Component}

The hot, X-ray emitting gas is assumed to be in hydrostatic equilibrium
so that
\begin{equation}
{d\over {dr}}\left({{{\rho_{\rm gas}kT}\over{\mu m_p}}}\right)=
-{{GM(<r)}\over {r^2}}\rho_{\rm gas}.
\end{equation}
The mass in hot gas
is not significant compared to the stellar mass, so its contribution
to the gravitational potential is neglected.
Since we are interested in the average
properties of the DW sample, and because spatial
analysis is not available for most of the sample, we assume
that ${\rho_{\rm gas}}^2\propto \rho_*$; this has been found to be a
good approximation for X-ray bright galaxies,
and is expected if heating is balanced by cooling locally (\cite{lm87}).
Both assumptions, hydrostatic equilibrium and
${\rho_{\rm gas}}^2\propto \rho_*$, are more likely to break down
for galaxies with low X-ray-to-optical luminosity ratios, where a
partial or transient outflow might obtain; however, the
$T$--$\sigma$ relation for these relatively X-ray faint
systems does not differ, within the statistics of the sample, from
that of the sample as a whole.

\subsection{Scaling Relations and Model Solutions}

A significant unifying simplification for the
HST stellar models follows from the observations that
the ratios of core mass to total stellar mass ($M_{\rm core}/M_*(r_{\rm max})$,
where $M_{\rm core}$ is $L_{\rm core}$ [equation 4] multiplied by the
stellar mass-to-light ratio), and
break radius $r_{\rm br}$ to $r_e$ (the ``true'' $r_e$
as observed, not the half-light radius in our models) are approximately
constant: $M_{\rm core}\approx 0.012M_*(r_{\rm max})$ and $r_{\rm br}\approx r_e/33$
(\cite{f97}).
Our adoption of these scaling relations, along with the empirical
relation $M_*/L_V\propto {L_V}^{1/4}$, insures that our model galaxies are
consistent with the fundamental plane (FP) as long as the core is 
baryon-dominated, since it follows that
\begin{equation}
{\sigma_c}^2\propto {{M_{\rm core}}\over {r_{\rm br}}}
\propto {{M_{*}}\over {r_e}}\rightarrow L_V\propto{\sigma_c}^{8/5}{r_e}^{4/5},
\end{equation}
where $\sigma_c$ is the central
velocity dispersion.\footnote{We note at this point that we do not 
distinguish between
``dark'' (``luminous'') and ``non-baryonic'' (``baryonic'') matter.
Elliptical galaxies may very well have an additional {\it baryonic}
dark component in the form of stellar remnants that may 
account for the variation with luminosity of the central mass-to-light
ratio (\cite{zs96}). Any such contribution is subsumed into our definition of
the stellar (``luminous'') component through $M_*/L_V$.}
In order to provide us with a
one-parameter family of models
(the quality and quantity of the X-ray data is insufficient
to attempt to derive 
a ``$T$--$\sigma$--$r_e$'' or equivalent FP-like relation),
we take a ``Faber-Jackson cut'' through the FP such that
$L_V\propto \sigma^4\rightarrow r_e\propto L_V^{3/4}$, normalized
so that $r_e=6.3$ kpc (and $\sigma=250$ km s$^{-1}$)
at $M_V=-22$
for $H_0=80$ km s$^{-1}$ Mpc$^{-1}$. 
The root-mean-square deviation of galaxies in the DW sample
from this projection of the FP is only 34 km s$^{-1}$
(13.5\% of the mean).
The Hernquist models (equation 5) likewise
adhere to the FP, since we universally adopt $r_{\rm Hern}=0.45r_e$.

Equations (6) and (10) are solved in dimensionless form. There are 
two fundamental
(luminosity-dependent) dimensional quantities 
that we denote as $r_*$ (chosen to be $r_{\rm br}$ for
the HST model or $r_{\rm Hern}$ for the Hernquist model)
and $M_0$ ($M_{\rm core}$ for
the HST model or total stellar mass $M_*$ for the Hernquist model).
Thus, adopting the
above scaling relations,
the length-, mass-, density-, velocity-, and temperature-scaling relations
are as follows:
\begin{equation}
r_*\propto\lambda^{3/4}{h_{80}}^{-1},
\end{equation}
\begin{equation}
M_0\propto\lambda^{5/4}\mu_*{h_{80}}^{-1},
\end{equation}
\begin{equation}
\rho_0\equiv {{M_0}\over {4\pi{r_*}^3}}
\propto\lambda^{-1}\mu_*{h_{80}}^{2},
\end{equation}
\begin{equation}
\sigma_0\equiv \left({{GM_0}\over r}\right)^{1/2}
\propto\lambda^{1/4}{\mu_*}^{1/2},
\end{equation}
and
\begin{equation}
kT_0\equiv \mu m_p{\sigma_0}^2
\propto\lambda^{1/2}\mu_*.
\end{equation}
Here $h_{80}$ is the Hubble constant in units of
80 km s$^{-1}$ Mpc$^{-1}$, $\lambda\equiv L_V/L_0$ with
$L_0=5.2\times 10^{10}{h_{80}}^{-2}$L$_{V_\odot}$,
and $\mu_*$ is defined such that the stellar mass-to-light ratio
$M_*/L_V=10\mu_*$M$_{\odot}/$L$_{V_\odot}$ for $L=L_0$ and $h_{80}=1$.
$L_0{h_{80}}^2$ corresponds to $M_V=-22$, {\it i.e.} $L_0\approx 3L_*$.

The dark matter mass distribution is scaled accordingly, {\it i.e.}
\begin{equation}
{{M_{\rm dm}}\over {M_0}}={\alpha}
{{f(x/\delta)}\over {f(x_{\rm max}/\delta)}}
{{M_*(r_{\rm max})}\over {M_0}},
\end{equation}
where $\alpha$ is the ratio of dark to stellar mass at the
HST stellar model cutoff radius, i.e.
\begin{equation}
\alpha \equiv \left. {M_{\rm dm} \over M_*}\right|_{r_{\rm max}},
\end{equation}
and $\delta\equiv r_{\rm dm}/r_*$ is the dimensionless
dark matter scale length. In the above, $x\equiv r/r_*$,
$x_{\rm max}\equiv r_{\rm max}/r_*$, and the function $f$ is defined
in equation (9).

To determine the average values of $T$ and $\sigma$,
the hydrostatic and Jeans equations are integrated inward from a
zero-pressure boundary at $r=\infty$
(see \cite{l94}). Although the
stellar mass has a cutoff ($r_{\rm max}$)
for the HST model, the gas and dark matter halos
are likely to extend beyond this and are not truncated in the models.
This results in a pressure contribution to the gas virial
temperature within $r_{\rm max}$ that leads to
lower values of $\beta_{\rm spec}$ for a given $\alpha$, and is therefore
conservative in the sense of minimizing the required amount of
dark matter.

The average quantities that constitute the observable
$\beta_{\rm spec}$ are calculated as follows:
we define $\langle\sigma\rangle$ as the average projected
stellar velocity dispersion within a circular aperture of size
$r_{\rm br}$, and $\langle T\rangle$ as the emission-weighted projected
gas temperature within $r_{\rm max}$ (the same aperture
over which the temperature is measured in DW, corresponding to
the radius enclosing $\approx 90$\% of the optical light). 
The value of $\beta_{\rm spec}$ is robust
to the precise
choice of these radii since $r_{\rm max}\gg r_e \gg r_{\rm br}$.
The aperture-averaged,
projected velocity dispersion for all central
radii $>0.1r_{\rm br}$ ($>0.3r_{\rm br}$) 
is within 10\% of its value at 
$r_{\rm br}$ for the HST (Hernquist) model,
assuming isotropic orbits and that the central potential
is dominated by the stellar component.
By
averaging over constant multiples of the fundamental stellar
scale length ($r_*$), we assure that our computed values of $\beta_{\rm spec}$ 
depend only on the {\it relative} (to stellar) dimensionless 
dark matter and velocity
dispersion anisotropy parameters.
These parameters are the following: 
the ratio of dark to luminous matter scale length,
$\delta\equiv r_{\rm dm}/r_*$; the mass ratio $M_{\rm dm}/M_*$ within $r_{\rm max}$,
$\alpha$; and the dimensionless
anisotropy scale length, $s/r_*$. 
The only additional model parameters are the stellar distribution slopes.
For a given triplet
($s/r_*$, $[M_{\rm dm}/M_*]_{r_{\rm max}}$, $r_{\rm dm}/r_*$),
a general solution for
the dimensionless gas temperature and stellar velocity dispersion
radial distributions follows that can be
scaled to physical units for any $h_{80}$ and $L_V$
using equations (12--16); $\mu_*$ is determined by
normalizing to $\langle\sigma\rangle=250$ km s$^{-1}$
for $M_V=-22$ and $h_{80}=1$;
note that $\mu_*\propto(\sigma_0/\langle\sigma\rangle)^2$.

\section{Results}

In this section we consider
the dependences of $\beta_{\rm spec}$ on
$s/r_*$, $r_{\rm dm}/r_*$, and $[M_{\rm dm}/M_*]_{r_{\rm max}}$
for both HST and Hernquist models. While the latter provides a more
accurate global representation
of elliptical galaxy stellar distributions, the former is more accurate 
around and inside
the break radius $r_{\rm br}$ where $\langle\sigma\rangle$ is calculated. For 
models that are dark matter dominated beyond several $r_e$, the 
redistribution
of the stellar mass at large radii in
the HST model is irrelevant.

\subsection{Variation with Anisotropy Scale \& the Necessity for Dark Matter}

In the absence of dark matter, the only free parameter in our models
is the anisotropy length scale $s$; recall that this is defined such that
orbits become isotropic for $r\ll s$ and radial for $r\gg s$.
Therefore, if stars dominate the gravitational potential within
$r_{\rm max}$, and if the shape of the velocity 
ellipsoid does not vary 
systematically with $L_V$,
then $\beta_{\rm spec}$ should be independent of
$L_V$ and $\langle\sigma\rangle$
({\it i.e.},
$\langle T\rangle\propto {\langle\sigma\rangle}^{2}$), 
contrary to the observed $T$--$\sigma$ relation (DW).
Moreover, as we now show, the typical observed value of 
$\beta_{\rm spec}\approx 0.5$
cannot be reproduced in the absence of dark matter.

The solid and dotted curves in Figure 3 show the variation
of $\beta_{\rm spec}$ with the 
dimensionless anisotropy scale length $s/r_{\rm br}$ for, respectively,
the HST and Hernquist stellar models without dark matter.
Evidently, $\beta_{\rm spec}$ falls with increasing $s$,
but is essentially constant for
$s>10$--$20r_{\rm br}$ ($\approx r_e/3$--$2r_e/3$), since $\langle\sigma\rangle$
is an average over a much smaller radius. The asymptotic values
of $\beta_{\rm spec}$ are $1.2$ and $0.75$ for the 
the HST and Hernquist stellar models, respectively.
The lower value for the Hernquist model is the result of a temperature
maximum at several $r_{\rm br}$ that substantially contributes
to the emission-averaged temperature because of the high stellar
(and, by assumption, gas) density there. As dark matter is introduced,
the temperature profile becomes more nearly
isothermal, and differences in $\beta_{\rm spec}$
between HST and Hernquist models lessen (see next paragraph, and Figure 3). 
Varying the slope parameters of the HST models (equation 2)
does not result in any values of $\beta_{\rm spec}$ less than 0.75.

The values of $\beta_{\rm spec}$ derived without dark matter lie
well above what is observed in a typical elliptical
galaxy (in fact, no galaxy in the DW sample has a best-fit value
of $\beta_{\rm spec}$ greater than 1; see Figure 1).
One of our major conclusions is already apparent. While
a typical bright elliptical galaxy has $\beta_{\rm spec}\approx 0.5$,
{\it even under the most extreme assumptions,
$\beta_{\rm spec}$ is never less than $\approx 0.75$ without the presence of
an extended dark halo}. While 10 of the 30 galaxies in the DW sample have
upper limits to $\beta_{\rm spec}$ greater than 0.75, only 3 have 
upper limits greater than 1.2 (see Figure 1). 
Of the 16 galaxies where $\beta_{\rm spec}$
is determined to better than 20\%, only 1 has an upper limit greater than 0.75.

The dash-dotted and dashed curves show the corresponding models
containing dark matter halos with dark
matter scale length $r_{\rm dm}$ set equal to
$r_e$ ({\it i.e.} $r_{\rm dm}=33r_{\rm br}=2.2r_{\rm Hern}$), and with 
$[M_{\rm dm}=3M_*]_{r_{\rm max}}$ 
chosen so that $\beta_{\rm spec}\approx 0.5$ for $s\ge r_e$.
The non-luminous
fraction of the total mass within $r_e$
is $\approx 0.6$ and $\approx 0.5$ for the HST and
Hernquist models, respectively.

From here on we assume isotropic stellar orbits. As shown in Figure 3,
for the practical purposes of calculating $\beta_{\rm spec}$, this is
equivalent to assuming that the orbits are mostly isotropic inside
$\sim r_e/2$. Since $\beta_{\rm spec}$ increases with decreasing
$s/r_{\rm br}$, more dark matter is required to obtain a given value 
of $\beta_{\rm spec}$ if this assumption is discarded ---
once again we adopt the most conservative
assumption as regards the required amount of dark matter.

\subsection{Variation with Dark Matter Scale}

Figure 4a shows $\beta_{\rm spec}$ as a function of
the ratio of dark matter scale length
to stellar break radius, $r_{\rm dm}/r_{\rm br}$,
for the HST (solid curve) and Hernquist (dotted curve) 
models with $[M_{\rm dm}=3M_*]_{r_{\rm max}}$. (For the Hernquist model,
the scaling relation $r_{\rm br}=r_e/33=r_{\rm Hern}/15$ has been utilized.)
As $r_{\rm dm}$ decreases below $\sim 10r_{\rm br}$, dark matter plays an increasingly
important role in determining $\langle\sigma\rangle$,
so $\beta_{\rm spec}$ rises.
For fixed $[M_{\rm dm}/M_*]_{r_{\rm max}}$, $\beta_{\rm spec}$ is nearly 
independent of $r_{\rm dm}/r_{\rm br}$
provided that $r_{\rm dm}>r_e$. It is clear that measurements of
$\beta_{\rm spec}$ can only place {\it lower} limits on $r_{\rm dm}$.

There is a fair
amount of freedom in how the dark matter can be distributed and
still produce $\beta_{\rm spec}\approx 0.5$.
This is illustrated in Figure 4b,
which displays the variation in baryon fraction at $r_e$ (solid curve
for the HST stellar model, dotted curve for the Hernquist model) and at
$r_{\rm br}$ (dashed curve
for the HST stellar model, dot-dashed curve for the Hernquist model).
$\beta_{\rm spec}$ changes very slowly with $r_{\rm dm}/r_{\rm br}$
for $r_{\rm dm}/r_{\rm br}>20$, while the baryon fraction inside
$r_{\rm br}$ varies from $\approx 0.73$ and $\approx 0.66$ for the HST and
Hernquist stellar
models, respectively, to $\approx 0.98$ for both models as $r_{\rm dm}$ becomes
very large.
Meanwhile, the baryon fraction inside $r_e=33r_{\rm br}$
increases from $\approx 0.31$ ($\approx 0.43$) 
if $r_{\rm dm}/r_{\rm br}=20$, to $\approx 0.75$ ($\approx 0.84$) 
for the HST (Hernquist) model as
$r_{\rm dm}$ becomes large compared to the stellar cutoff radius $r_{\rm max}$.

\subsection{Variations with Dark-to-Luminous Mass Fraction}

Figure 5 shows $\beta_{\rm spec}$ as a function of 
$[M_{\rm dm}/M_*]_{r_{\rm max}}$
for HST and Hernquist stellar models
with $r_{\rm dm}$ set equal to
0.5, 1.0, and $2.0r_e$.
As discussed above,
$\beta_{\rm spec}\rightarrow 1.2$ (0.75) 
as $M_{\rm dm}\rightarrow 0$ for the HST (Hernquist) model.
The curves pass through $\beta_{\rm spec}\approx 0.5$
at $[M_{\rm dm}\approx 3M_*]_{r_{\rm max}}$
if $r_{\rm dm}=r_e$
or $2r_e$, while 
$[M_{\rm dm}\approx 4(6)M_*]_{r_{\rm max}}$ is
required if $r_{\rm dm}=r_e/2$ for HST (Hernquist) models.
As $[M_{\rm dm}/M_*]_{r_{\rm max}}$ becomes sufficiently large
for fixed $r_{\rm dm}$, dark matter comes to dominate the mass
even at small radii ({\it i.e} within $r_{\rm br}$ where $\langle\sigma\rangle$
is calculated) and $\beta_{\rm spec}$ becomes independent of $\alpha$.
That is, in order to maintain
consistency with the FP and the Faber-Jackson relation,
the physical mass must remain constant and the increase in dark-to-luminous
mass ratio $\alpha$
is compensated for by decreasing the stellar mass-to-light
ratio ({\it i.e.}, $\mu_*$ in equations 13--16) accordingly.

\subsection{Limits on Dark Matter Parameters}

We have shown (Section 4.1) that the lower limits on
$\beta_{\rm spec}$ for the HST and Hernquist stellar models
without dark matter are 0.75 and 1.2, respectively.
To explain observed values of $\beta_{\rm spec}$ that are less than this
requires the addition of dark matter halos with sufficient mass
to raise the value of
$\langle T\rangle$, but with a large enough scale length
to leave $\langle\sigma\rangle$ (which is averaged over a much
smaller scale) relatively unaltered. By calculating solutions
to equations (6) and (10) for the entire 
($[M_{\rm dm}/M_*]_{r_{\rm max}}, r_{\rm dm}/r_*$)
parameter space, we have derived limits on the
amount and scale length of any dark halo able to
reproduce a given observed value of $\beta_{\rm spec}$.

As discussed in Section 4.3 and shown in Figure 5, for a fixed
value of
$r_{\rm dm}/r_{\rm br}$
there is a minimum value of $\beta_{\rm spec}$
that is obtained when $\alpha$ becomes sufficiently large that
both $\langle\sigma\rangle$ and $\langle T\rangle$ are
determined by the non-luminous mass component. Since this minimum
$\beta_{\rm spec}$ increases as $r_{\rm dm}/r_{\rm br}$ decreases, there is a minimum
value of $r_{\rm dm}/r_{\rm br}$ required to explain any observed $\beta_{\rm spec}$.
Figure 6a shows these minimum values of $r_{\rm dm}/r_{\rm br}$ for
the HST (solid curve) and Hernquist (dotted curve) models. The
minimum value of $r_{\rm dm}$ for a
$L_V\approx L_0\approx 3L_*$
galaxy with $\beta_{\rm spec}\approx 0.5$ is $\approx 2{h_{80}}^{-1}$ kpc
($\approx r_e/3$); the dimensional value
scales as ${L_V}^{3/4}$.
These extreme models are dark matter dominated,
even inside the break radius $r_{\rm br}$, and have very
low stellar mass-to-light ratios. Requiring models to have
$M_*/L_V\ge 5$M$_{\odot}/$L$_{V_\odot}$ for $L_V=L_0$ and $h_{80}=1$
(or, equivalently, that $\gtrsim$60\% of the mass within
$r_{\rm br}$ be stellar)
roughly
doubles the minimum values of $r_{\rm dm}/r_{\rm br}$ for
$0.4<\beta_{\rm spec}<0.6$.
This reasonable requirement then implies that
$r_{\rm dm} > 4(L_V/3L_*)^{3/4}h^{-1}_{80}$ kpc.

The maximum baryon fractions within $r_{\rm br}$, $r_e$,
and $r_{\rm max}$ for
any given $\beta_{\rm spec}$.
are shown in Figure 6b.
The lower limits on the dark matter fractions
in models required to explain $\beta_{\rm spec}>0.4$ are less
than 5\% within $r_{\rm br}$, and less than 20\% within $r_e$ ---
non-luminous halos could be quite inconspicuous 
unless data well beyond $r_e$ are considered.
The maximum baryon fractions at $r_{\rm max}$ for the
HST (Hernquist) model are 0.43 (0.15), 0.53 (0.27), and
0.61 (0.45) to produce
$\beta_{\rm spec}=0.4$, 0.5, and 0.6, respectively.
These models have very large  dark matter scale lengths
($r_{\rm dm}\gg r_{\rm max}$) and are analogous to ``maximum disk'' models of
spiral galaxies in that they maximize the contribution of stars
to the mass-to-light ratio.

\subsection{Scaling of Dark Matter Content with Optical Luminosity}

In order to investigate the scaling characteristics
of dark halos with optical luminosity, we consider the
equivalent of the $T$--$\sigma$ relation in the $\beta_{\rm spec}$--$L_V$
plane. Correlations of $\beta_{\rm spec}$ and log$(\beta_{\rm spec})$
with log($L_V$) are derived using the DW data set, blue
magnitudes and $B-V$ colors from the RC3 catalog (\cite{rc3}), and velocity
distances from Faber et al. (1989). (The 
$\beta_{\rm spec}$--log($L_V$) data and best-fit correlation 
are re-plotted in Figure 7b.)
Galaxies in the centers of cluster cooling flows and
those dominated by nuclear point source emission are excluded. 
The log$(\beta_{\rm spec})$--log($L_V$) correlation, equivalent to
$\beta_{\rm spec}\approx 0.58\lambda^{0.13}$ where 
$\lambda$ is the visual luminosity in units of
$L_0=5.2\times 10^{10}{h_{80}}^{-2}$L$_{V_\odot}$, is
essentially identical to the convolution of the $T$--$\sigma$ 
correlation and Faber-Jackson relation (see Section 3.4).
Nevertheless, requiring that
models conform to the observations in this form makes use of the
tilt of the FP for scaling of the
the stellar component, thus circumventing concerns about
the thickness of the projection of the FP
onto the $L_V$--$\sigma$ plane.
Note that the galaxies considered span $0.15<\lambda<3$, although most of the
sample lies within the range $0.5<\lambda<2$.

Given our homologous elliptical
galaxy models, which are
based on the scaling properties of the FP, the
observed variation in $\beta_{\rm spec}$ implies
that there is a systematic variation of
relative dark matter properties which breaks the
self-similarity for galaxies of
different luminosity.
The sense of the observed $T$--$\sigma$ relation could also
be explained by
a fine-tuned systematic variation of the velocity dispersion
anisotropy (having more radial orbits for more luminous galaxies), or if
more luminous galaxies have cuspier
stellar distributions (larger values of $c$ in equation 2).
However, the latter is
contrary to observations; and as we have shown,
the observed values of $\beta_{\rm spec}$ are too low to be explained
without the introduction of dark matter.
It is possible to explain the $\beta_{\rm spec}$--$L_V$
trend,
assuming the dark-to-luminous mass ratio within $r_{\rm max}$ is constant,
if the ratio of dark matter to stellar break (or, equivalently, effective) 
radius, $r_{\rm dm}/r_{\rm br}$
decreases with increasing galaxy luminosity $L_V$.
However, since $\beta_{\rm spec}$ is flat for
$r_{\rm dm}/r_{\rm br}>10$ and steeply rises for $r_{\rm dm}/r_{\rm br}<10$ (Figure
4a) this requires fine-tuning. Moreover, as we
discuss shortly,
$r_{\rm dm}/r_{\rm br}<10$ is not expected.

More naturally, the increase in $\beta_{\rm spec}$ with $L_V$ could
be explained if $[M_{\rm dm}/M_*]_{r_{\rm max}}$
{\it decreases} with {\it increasing} $L_V$ (that is, if more luminous
galaxies are less dark matter dominated)
and $r_{\rm dm}/r_{\rm br}$ is relatively large (see Figure 5). We now discuss the
plausibility of such a correlation within the context of the
hierarchical clustering formalism of structure formation, simulations
of which produced the dark matter distribution expressed by equations (8)
and (9). This formalism provides specific predictions for $r_{\rm dm}$
as a function of virial mass (defined here as the mass within a radius
where the mean dark matter
density is 200 times the average density in the universe)
for a given cosmology and power spectrum
of initial fluctuations.

In order to extend our models to the scale of the virial radius and mass,
we transform the 
dark matter parameters $r_{\rm dm}/r_{*}$ and $[M_{\rm dm}/M_*]_{r_{\rm max}}$
that are directly constrained by observations into the
following pair of global parameters:
the dark matter concentration parameter 
\begin{equation}
C\equiv r_{200}/r_{\rm dm}, 
\end{equation}
and the global dark-to-luminous mass ratio 
\begin{equation}
\alpha_{\rm virial}\equiv [M_{\rm dm}/M_*]_{r_{200}}, 
\end{equation}
where $r_{200}$ refers
to the radius within which the overdensity of the dark matter (only) is 200.
($M_{\rm dm}(r_{200})$ and $r_{200}$ refer 
to the original mass of the dark halo, but
this may have been subsequently reduced in rich environments
through tidal stripping.)
Note that for $C=10$ and $\alpha_{\rm virial}=10$, 
$r_{200}=320{h_{80}}^{-1}$ kpc, $r_{\rm dm}/r_{\rm br}=170$, and 
$[M_{\rm dm}/M_*]_{r_{\rm max}}=1.6$ for $\lambda=1$ ($L\approx 3L_*$).

In bottom-up hierarchical clustering theories of structure formation,
lower mass halos are more concentrated --- a reflection of the higher density
in the universe at the earlier epoch of their formation (\cite{nfw}). 
We parameterize this dependence with the function
\begin{equation}
C={{C_o}\over {\chi^{\gamma}+B/\chi}}
\end{equation}
for the concentration,
where $\chi\equiv M_{\rm dm}(r_{200})/M_o$ and 
$M_o\equiv 10^{13}{h_{80}}^{-1}$M$_{\odot}$ is on the order of the 
present-day characteristic mass. The second term in the denominator of
equation (21) is introduced to prevent the concentration from
exceeding a maximum value $C_{\rm max}$, that we set equal to 100,
and driving the stellar
mass-to-light ratio to very small values. Note that
$B=[C_{\rm max}(1+\gamma)/C_o]^{-(1+\gamma)/\gamma}$ becomes insignificant
for $\gamma\ll 1$ (e.g, in standard CDM models).

Utilizing
the ``characteristic density''
program kindly provided by
J. Navarro, a standard cold dark matter (CDM) fluctuation spectrum
results in $C_o\approx 10$ and $\gamma\approx 0.1$.
For these parameters, 
$r_{\rm dm}/r_{\rm br}\approx 150(\alpha_{\rm virial}/10)^{13/30}\lambda^{-5/24}$
($r_{\rm dm}/r_e\approx 4.5$ for $\alpha_{\rm virial}\approx 10$ and 
$\lambda\approx 1$) 
and $r_{\rm dm}$ is indeed much greater than $10r_{\rm br}$ in the luminosity range
of interest.
The dot-dashed curve in Figure 7a shows $[M_{\rm dm}/M_*]_{r_{\rm max}}$
as a function of $\lambda$
for $\alpha_{\rm virial}=16$ ({\it i.e.} baryon fraction 0.06)
CDM models
(and for two alternative models to be discussed
in the next section).
Clearly $[M_{\rm dm}/M_*]_{r_{\rm max}}$
{\it increases} with optical
luminosity, implying
a {\it decrease} of
$\beta_{\rm spec}$ with increasing
$L_V$ as shown by the dot-dashed curve in Figure 7b.
A comparison with the observed 
$\beta_{\rm spec}$--$\lambda$ correlation
(solid curve in Figure 7b), demonstrates that
{\it a constant baryon fraction CDM model for elliptical galaxy
dark matter halos
clearly fails to match the observed correlation}, even over the limited
range of $\lambda$ where the correlation is
well-established. A successful alternative model
must produce dark halos where $[M_{\rm dm}/M_*]_{r_{\rm max}}$ {\it decreases}
with increasing optical
luminosity in this range. We will discuss the form such
possible alternatives must take in the next
section.

\section{Discussion}

A natural explanation for why $\beta_{\rm spec}$,
the ratio of the central stellar velocity
dispersion to the globally-averaged gas temperature,
{\it increases} with
$\langle\sigma\rangle$ and therefore $L_V$ is that the
dark-to-luminous mass
ratio within the luminous parts of elliptical galaxies ($r<r_{\rm max}=6r_e$)
{\it decreases} with $L_V$ (see Figures 5, 7a and 7b). 
In fact, the observed trend requires that
the total mass-to-light ratio within $r_{\rm max}$
be nearly constant.
This can be understood from a simple argument based on
the virial theorem and FP, as follows. The virial temperature
within $r_{\rm max}$, 
$T_{\rm virial}\propto (M_{\rm dm}+M_*)/r_{\rm max}
\propto (1+M_{\rm dm}/M_*)(M_*/r_e)$ since $r_{\rm max}\propto r_e$ (all masses
are evaluated at $r_{\rm max}$). Since, from the FP
$\langle\sigma\rangle^2\propto M_*/r_e$, 
it follows that $\langle\sigma\rangle^2/T_{\rm virial} \propto
(1+M_{\rm dm}/M_*)^{-1}$, or $\beta_{\rm spec}\propto
(1+M_{\rm dm}/M_*)^{-1}$ if we 
associate $T_{\rm virial}$ with the integrated
X-ray temperature, which should be a fair approximation.

The implication from the observed $\beta_{\rm spec}$--$L_V$
trend that the total mass-to-light
ratio within $r_{\rm max}$ be nearly constant (see below) is
contrary to expectations from
CDM models of large scale structure
if all elliptical galaxies have the same global baryon fraction. 
The dark-to-luminous mass ratio is predicted to {\it increase} with $L_V$,
in such models,
since the ratio
of the dark matter scale length $r_{\rm dm}$ to
the optical half-light radius is large and weakly varying, thus
producing a predicted
correlation between $\beta_{\rm spec}$ and $L_V$ in the opposite
sense to what is observed
(Figures 7a and 7b). We consider two simple variations on the 
above model to try and recover the observed $\beta_{\rm spec}$--$L_V$ trend.
First we assume that the dark matter concentration depends only weakly 
on halo mass, as CDM predicts, but relax the 
constant baryon fraction assumption. Then we assume that the concentration
varies much more steeply with virial mass than the CDM prediction.

\subsection{Alternatives to Constant Baryon Fraction CDM}

If the weak dependence of concentration on halo mass predicted in
CDM simulations (equation 21 with $C_o=10$ and $\gamma=0.1$) is maintained,
then the assumption of a constant baryon fraction within $r_{200}$ must 
be relaxed: the required increase in dark matter fraction 
within $r_{\rm max}$ for
less luminous galaxies must be a reflection of a larger
{\it total} ({\it i.e.} within $r_{200}$)
dark-to-luminous mass ratio. If all
galaxies start off with identical baryon fractions, this then implies that
smaller galaxies lose an increasingly large fraction of their original
baryonic mass. (Similar conclusions for spiral galaxies have
been reached by \cite{pss}.)
We parameterize this variation in the dark/stellar mass ratio as
\begin{equation}
\alpha_{\rm virial}=
\alpha_{\rm min}\left[1+\left({{\lambda}\over {\lambda_o}}\right)^{-\nu}\right],
\end{equation}
where, as previously defined, 
$\alpha_{\rm virial}\equiv [M_{\rm dm}/M_*]_{r_{200}}$,
$\lambda$ is the visual luminosity in units of
$L_0=5.2\times 10^{10}{h_{80}}^{-2}$L$_{V_\odot}$, and $\lambda_o$ is
a fitting parameter. 
Equation (22) has the properties $\alpha_{\rm virial}\rightarrow\infty$
as $\lambda\rightarrow 0$ (complete baryonic mass loss), and
$\alpha_{\rm virial}\rightarrow\alpha_{\rm min}$
as $\lambda\rightarrow\infty$ (presumably, $(1+\alpha_{\rm min})^{-1}$ is
the average baryon fraction in the universe). The observed 
$\beta_{\rm spec}$--$L_V$ correlation is 
satisfactorally reproduced in models with $\nu=5/4$, 
${\lambda_o}=3$, and $\alpha_{\rm min}=0.5$--2.5.
The dotted curves in Figures 7a and 7b represent HST stellar models 
with $\alpha_{\rm min}=1.5$; the observed trend is
very well matched (Figures 7b). (Hernquist stellar models require
even larger values of $\nu$; however, this is an artifact of the
underestimate of $\beta_{\rm spec}$ for low $M_{\rm dm}$; see section 4.1).

Alternatively, if the dark matter halo concentration
increases much more steeply with
virial mass than predicted for a CDM fluctuation spectrum,
the observed $\beta_{\rm spec}$--$L_V$ correlation can be reproduced assuming
a constant total baryon fraction ({\it i.e.}, $\alpha_{\rm virial}$). 
In this case, the necessary decrease in $[M_{\rm dm}/M_*]_{r_{\rm max}}$
for more luminous galaxies results not from a smaller 
relative global amount of dark matter, but from a larger fraction of the dark
matter lying outside of the luminous portion of the galaxy ({\it i.e.}
$r_{\rm max}$). We have found that equation (21) with $\gamma=1$,
$C_o=4$, and $\alpha_{\rm virial}>5$ provides models that 
adequately reproduce the data (smaller values of $\gamma$ require
larger values of $\alpha_{\rm virial}$).
$[M_{\rm dm}/M_*]_{r_{\rm max}}$ and $\beta_{\rm spec}$
are shown as functions of $L_V$, for $\alpha_{\rm virial}=16$,
by the dashed curves in Figure 7a and 7b.
Such a steep dependence of the concentration on total mass
would seem to indicate a flatter-than-CDM 
primordial fluctuation spectrum (\cite{nfw}).

These alternative scenarios were designed to reproduce the
$\beta_{\rm spec}$--$L_V$ correlation
without consideration of the scatter or that of galaxies about
the mean FP. We address these concerns
by computing $\beta_{spec}$ for individual pairs $(L_V$, $\sigma)$ 
corresponding to each of the 25 non-cluster-cooling-flow, 
non-point-source-dominated 
galaxies in the DW sample, for the three scenarios of Figures 7a--b.
A comparison with the observed values
of $\beta_{\rm spec}$, as shown in Figures 8a--c, confirms that the 
constant baryon fraction CDM scenario provides the poorest match to the
observations, on average, and further
illustrates the magnitude of variations
about the mean correlation. Further investigation into these
variations will be pursued in a future work that examines the individual
galaxies in detail.

Both of these alternative scenarios have drawbacks. 
Mass loss induced by supernovae-driven galactic winds during the
star-forming epoch of elliptical galaxies has been inferred
from both the color-magnitude diagram and the enrichment
of the intracluster medium (e.g., \cite{lm96}, \cite{g97}).
However the magnitude of this effect must be extreme 
({\it i.e} $\nu$, in equation 22, on the order of 1
or greater) if it is to explain the
increasing dominance of dark matter in less luminous galaxies required
to explain the observed $T$--$\sigma$ relation. The mass loss
implied by equation (22) with the required set of parameters, when
integrated over the elliptical galaxy luminosity function (\cite{mhg};
\cite{l96}), corresponds to $\sim 80$\% of the original baryonic
mass of $L>L_*$ galaxies. Moreover, in such a scenario, all
elliptical galaxies with $L_V<3\times 10^{11}h_{80}$L$_{V_\odot}$ 
would span a range in total mass of less than a factor of three; the
large range in mass loss could not, therefore, be due to less luminous
galaxies lying within shallower potential wells. Instead, one must presume
that variations in the relative timing of
star formation and the merging of pre-galactic fragments is 
responsible for the large range in mass loss, with galaxies with smaller
present-day luminosities having lost the largest fraction of their original
baryonic mass. Light would be a very poor tracer of mass 
for elliptical galaxies, and
the luminosity function would be a reflection of
the rarity of elliptical galaxies sufficiently relaxed at
the star formation epoch to prevent copious mass loss
from galactic winds. 

It might therefore seem more appealing to resort to a concentration-mass
relationship for dark halos (equation 21) that is steeper than 
predicted in CDM simulations
in such a way that optically less luminous galaxies are more
dark matter dominated
within $r_{\rm max}$, despite having the same total (within $r_{200}$) 
baryon fraction. The basic requirement is that the dark matter scale
length $r_{\rm dm}$ increase from $\sim r_e$ at the low-luminosity end of the
observed range, to $r_{\rm dm}\sim r_{\rm max}$ at 
$L_V\sim L_0=5.2\times 10^{10}{h_{80}}^{-2}$L$_{V_\odot}$, to
$r_{\rm dm}\sim r_{200}$ ($C\sim 1$) at the high luminosity range.
These criteria are met for only a rather narrow volume
in ($\alpha_{\rm virial}, C_o, \gamma$) parameter space, and thus this
scenario seems
to suffer from a fine-tuning problem; nonetheless, the range that
successfully reproduces the observed $T$--$\sigma$ relation
is not out of line for cosmogenic models with
flat primordial fluctuation spectra (\cite{nfw}).

\subsection{Mass-to-Light Ratios, Baryon Fractions \& Velocity Dispersions}

Figures 9a, 9b, and 9c show the luminosity dependences of 
mass, mass-to-light ratio, and baryon fraction at various radii
for the two models discussed above that reproduce
the observed $T$--$\sigma$ ($\beta_{\rm spec}$--$L_V$) relation:
dotted lines show the model with
CDM-predicted halo concentration as a function of mass
and increasing mass loss with decreasing luminosity
($C_o=10$ and $\gamma=0.1$ in equation 21, 
$\alpha_{\rm min}=1.5$, $\nu=5/4$, and $\lambda_o=3$ in equation 22);
dashed curves show the model with steeper-than-CDM concentration
as a function of mass and constant baryon fraction 
($C_o=4$ and $\gamma=1$ in equation 21, 
$\alpha_{\rm min}=16$ and $\lambda_o=0$ in equation 22).
We refer to these as ``mass loss'' and ``non-CDM'' scenarios, respectively.
The stellar model is assumed to follow equation (3),
and the luminosity range in the plots roughly corresponds to that of
the DW sample ($\approx 0.6$--$2L_*$). 
The radii of interest are
$r_e=6.3\lambda^{3/4}{h_{80}}^{-1}$ kpc, $r_{\rm br}=0.03r_e$, and
$r_{\rm max}=6r_e$; $r_{200}{h_{80}}$ 
increases from 270 to 400 kpc for the mass loss
scenario and $r_{200}\approx 370\lambda^{5/12} {h_{80}}^{-1}$ kpc for
the $\alpha_{\rm virial}=16$, non-CDM scenario, where
$\lambda\equiv L_V/L_0$ with
$L_0=5.2\times 10^{10}{h_{80}}^{-2}$L$_{V_\odot}\approx 3L_*$.

There are a number of robust properties that hold for any model
(including Hernquist models)
to successfully reproduce the $T$--$\sigma$ relation.
Since the models are constrained by the observed properties
of the stars and by the 
average temperature of hot gas in hydrostatic equilibrium  
within $r_{\rm max}=6r_e$, the inferred
quantities at $r_{\rm max}$ are nearly model-independent:
$M(r_{\rm max})/L_V\approx 25h_{80}$M$_{\odot}$/L$_{V_\odot}$,
$f_{\rm baryon}(r_{\rm max})\approx 0.35\lambda^{1/4}$. 
Note also that $M_*\approx 4.8\ 10^{11}\lambda^{5/4}{h_{80}}^{-1}$M$_{\odot}$
for any model where the central potential is dominated by the stellar
component.

These values agree strikingly well with 
the constraints from statistical studies of gravitational
lensing (\cite{k95}; \cite{bbs}; \cite{g96}). The solid lines in
Figure 9a shows the mass as a function of luminosity, evaluated
at $r_e$ and $r_{\rm max}$ for the
softened isothermal sphere (SIS) model that provides a best-fit to the
observed weak gravitational shear from \cite{g96}.
It is remarkable that the mass-to-light scaling derived
from the shear that is dominated by $\sim L_*$ galaxies at intermediate
redshifts is in such excellent agreement with our results using
a completely independent method
on local galaxies primarily with
$L>L_*$. For the constant baryon fraction CDM model, the total mass
within $r_{\rm max}$ is a significantly steeper function of luminosity
than is demonstrated by the curves in Figure 9a.

On scales both larger and smaller than $r_{\rm max}$,
the two scenarios depicted in Figures 9a-c diverge.
In the mass loss scenario
the baryon fraction within $r_e$ is generally greater than 0.7 and
$M(r_e)/L_V(r_e)\approx 12\lambda^{1/4}h_{80}$M$_{\odot}/$L$_{V_\odot}$,
while in the non-CDM scenario the mass-to-light ratio is
nearly constant, $M(r_e)/L_V(r_e)\approx 11h_{80}$M$_{\odot}/$L$_{V_\odot}$,
so that the
baryon fraction within $r_e$
becomes less than 0.5 below $L\approx 0.6L_*$. 

Extrapolated
out to the dark matter
virial radius, $r_{200}$, the total mass scalings
differ (by assumption). For the mass loss scenario,
\begin{equation}
M_{\rm virial}/L_V\approx 
9\left[A\lambda^{1/4}+B\lambda^{-1}\right]
h_{80}{\rm M}_{\odot}/{\rm L}_{V_\odot},
\end{equation}
where acceptable models have $A\approx 1.5$--5 and $B\approx 4$--8 
($A=2.5$, $B\approx 6$ for the model shown by dotted curves in
Figures 7a-b and 9a-c). 
For the non-CDM (or any
other constant baryon fraction) scenario,
\begin{equation}
M_{\rm virial}/L_V\approx 90(10f_{\rm baryon})^{-1}
\lambda^{1/4}h_{80}{\rm M}_{\odot}/{\rm L}_{V_\odot};
\end{equation}
the (constant) total baryon fraction
$f_{\rm baryon}\approx 0.06$ for the dashed curves shown in Figures 7a-b and 9a-c.
We found that 
$f_{\rm baryon}<0.15$ in non-CDM models is required to reproduce 
the observed $\beta_{\rm spec}$--$L_V$ correlation.
Non-CDM models with relatively 
small global $f_{\rm baryon}$ have larger baryon fractions
within $r_e$ for
$\lambda<1$ (because of larger dark matter scale lengths) and agree 
somewhat better with the lensing constraints at this radius.

Integrating over the early-type galaxy luminosity
function (\cite{mhg};
\cite{l96}) yields an average total ({\it i.e.} within the virial radius)
mass-to-light ratio $>65h_{80}$M$_{\odot}/$L$_{V_\odot}$
for the non-CDM (constant baryon fraction) scenario and
$\Omega_{\rm ellipticals}>0.02$, which can be
compared to
an average total
mass-to-light ratio of $\approx 150h_{80}$M$_{\odot}/$L$_{V_\odot}$
for spirals (\cite{pss}), and $\Omega_{\rm spirals}=0.04$--0.06.
In the mass loss scenario, the average total
mass-to-light ratio is sensitive to the lower luminosity
cutoff, $L_{\rm min}$, for the luminosity function and can be as high as
$\sim 1000h_{80}$M$_{\odot}/$L$_{V_\odot}$ 
($\Omega_{\rm ellipticals}\sim 0.4$) for 
$L_{\rm min}=10^8{h_{80}}^{-2}$L$_{V_\odot}$ since even the lowest
luminosity elliptical galaxy has a total mass of
$\sim 3\ 10^{12}{h_{80}}^{-1}$M$_{\odot}$. The average
density in the universe of virialized mass on galactic scales
is $>0.07$ of the critical density.

We have calculated
the velocity dispersion distributions for both the dark matter and
stellar density distributions, assuming isotropic orbits. 
These are compared in Figure 10 for an $L_V=L_0$ galaxy for both
the mass loss and non-CDM scenarios.
Both velocity dispersion profiles have maxima since the total gravitational
potential is not isothermal. The ratio 
(dark-matter-to-stars) of the squares of these maxima
is greater than 1.4
over the luminosity range in Figures 9a-c 
for the two models under consideration, and is $\approx 2$ over the range
$L_*<L_V<5L_*$.
In fact, the minimum value of this ratio for any HST or Hernquist model 
that produces $\beta_{\rm spec}<0.7$ is greater than one. In this sense, 
the dark matter is hotter than the stars, as predicted by Davis
\& White (1996).

\section{Summary and Conclusions}

In this work, we have addressed two essential features
of the X-ray temperatures derived by Davis \& White (1996)
for an optically complete sample of elliptical galaxies:
(1) the X-ray emitting
gas is always hotter than the stars and, typically, twice as hot
($\beta_{\rm spec}\equiv
{{\mu m_p\langle\sigma\rangle}^2/{k\langle T\rangle}}\approx 0.5$); 
(2) the gas/stellar temperature ratio tends to be higher for
galaxies with lower velocity dispersions 
(the ``$T$--$\sigma$ relation'').
We have constructed physically plausible models of
the mass distribution in bright elliptical galaxies
in an effort to constrain their average dark matter properties by
matching these observations.
The stellar models (described in \S3)
are fully consistent with the fundamental plane
scaling relations, and are
designed to either conform to the latest published {\it HST}
results on the structure of the centers of elliptical galaxies or
to follow the Hernquist (1990) approximation to a de Vaucouleurs profile
(see Figure 2).
We have made other plausible, but conservative,
assumptions, such as (1) maximizing the
non-gravitational ({\it i.e} pressure) contribution to the gas
temperature by allowing the gas and dark matter distributions to
extend to infinity and (2) assuming that the stellar orbits are
isotropic for $r\ll r_e$.

A basic and general result of our calculations is that,
in the absence of dark matter, $\beta_{\rm spec}\ge 0.75$ (see Figure 3). Since
$\beta_{\rm spec}\approx 0.5$ is observed, a convincing case is made that
dark matter is an extremely
common, if not ubiquitous, constituent of elliptical galaxies.
Although X-ray (and other) observations have been used to infer the presence
of dark matter in individual cases, we have shown that dark halos
are generic to luminous, nearby elliptical galaxies.
Furthermore, the dark matter/stellar temperature ratios (derived at
the peak of their velocity dispersion distributions, assuming 
isotropic orbits) are greater than one for models with
$\beta_{\rm spec}<0.7$. Thus, the observation that
the temperature of the extended hot gas exceeds the central stellar
temperature is a reflection of the fact that the 
dark matter is dynamically ``hotter"
than the stars, as suggested by Davis \& White (1996).

In \S4 we described how $\beta_{\rm spec}$ varies as functions of the two
relative parameters of the ``universal'' dark matter
distribution (equation 8) --- the ratio of dark matter to stellar
scale lengths, $\delta\equiv r_{\rm dm}/r_{*}$, and
the ratio of dark-to-luminous matter within $r_{\rm max}=6r_e$,
$\alpha\equiv [M_{\rm dm}/M_*]_{r_{\rm max}}$ (see Figures 4a-b and 5).
Because we attempt to match only the single
global observable $\beta_{\rm spec}$, there is an allowed range in the
details of the dark matter spatial distribution; however, we
have derived absolute limits
on the dark matter parameters required to obtain any particular value
of $\beta_{\rm spec}$ (see Figures 6a-b). 

The observations do not require that dark matter dominate the 
inner luminous regions of elliptical galaxies:
more than half of the mass within $r_e$ is 
baryonic for models with $\beta_{\rm spec}=0.5$ if $r_{\rm dm}>r_e$.

The most natural explanation of the tendency for galaxies with
lower stellar temperatures to have larger gas-to-stellar temperature
ratios
is that $[M_{\rm dm}/M_*]_{r_{\rm max}}$ decreases with $L_V$ in
such a way that, on average, the total mass-to-light ratio inside
$r_{\rm max}$ is nearly independent of optical luminosity. This ratio,
$\approx 25h_{80}$M$_{\odot}/$L$_{V_\odot}$, is exactly what
is predicted for mass models of elliptical galaxies designed to
explain the gravitational shear of background field galaxies.

If one specifies a scaling relation for the dark halo
concentration, one can
extend the dark matter distribution out to the virial radius and calculate the
total baryon fraction and mass-to-light ratio. When we attempt
to embed our models within the
CDM theory of hierarchical halo formation, the implied
dark matter
scaling badly fails to reproduce the observed $T$--$\sigma$ relation unless
smaller galaxies lose an increasingly larger fraction of their initial
baryonic content (see Figure 7b),
such that the average $L>L_*$ galaxy has lost
most of its initial baryonic mass.
Alternatively, the global dark-to-luminous mass ratio could be constant if
the dark halo concentration declines much more steeply with virial mass
than CDM models predict, so that the decrease in $[M_{\rm dm}/M_*]_{r_{\rm max}}$
for larger
systems is a result of a more diffuse dark matter halo
rather than a less massive (relative to the stars) halo. 
In this latter scenario, dark
matter may become increasingly important inside $r_e$ as $L_V$
decreases, becoming dominant for $L\ll L_*$. 
This deviation from CDM predictions of dark halo scaling
could conceivably be due to a relatively flat primordial fluctuation spectrum
on mass scales $<10^{14}$M$_{\odot}$
or to the effects on the
dark matter density profile of
the evolution of the baryonic component; to date, large
scale structure numerical simulations
that can resolve halos on galactic scales {\it and} include a
dissipational component have not been attempted.

In this paper we have shown that the observed relationship between
optical velocity dispersions and X-ray temperatures in giant
elliptical galaxies
implies that they have dark matter halos with
$M/L_V\approx 25h_{80}$M$_{\odot}/$L$_{V_\odot}$ within $6r_e$.
In the future, we plan to fully utilize the available
X-ray and optical
imaging and spectroscopic data for our sample on a case-by-case basis in
order derive the mass distributions in individual galaxies 
in the highest possible detail and
to investigate the scatter in the $T$--$\sigma$ relationship.
For galaxies with X-ray temperature profiles, we will be able
to constrain the detailed form of the dark matter distribution
as well as its integrated properties.

\acknowledgments

We thank Lars Hernquist for reminding us of the consistency considerations
of Ciotti \& Pelligrini (1992), and Richard Mushotzky for feedback on the
original manuscript.  Comments from an anonymous referee led to significant
improvements in the quality of this paper. R.E.W. acknowledges partial 
support from NASA grants NAG 5-1718 and NAG 5-1973.

\appendix

\section{Projection of a Broken Power-law}

Suppose the stellar density profile is given by a broken
power-law of the form
\begin{equation}
{\rho_{*}(r)}={\rho_{*o}}\left({r\over r_c}\right)^{-\kappa}
\left(1+{{r^2}\over{r_c^2}}\right)^{-\sigma}.
\end{equation}
The projected surface mass,
\begin{equation}
\Sigma_*(R)={\int_R}^{\infty}{{{\rho_{*}}d(r^2)}\over{\sqrt{r^2-R^2}}},
\end{equation}
can be recast, utilizing the change of variable
$y=(r^2-R^2)/R^2$,
into the form
\begin{equation}
\Sigma_*(b)={\rho_{*o}}r_cJ\left[ b^{-0.5(\kappa-1)}(1+b)^{-\sigma}\right],
\end{equation}
where
\begin{equation}
J(b)\equiv {\int_0}^{\infty}y^{-{1\over 2}}(y+1)^{-0.5\kappa}
(1+cy)^{-\sigma}dy,
\end{equation}
$b\equiv R^2/r_c^2$, and $c\equiv b(1+b)^{-1}$.
Note that the bracketed term
in equation (A3) has the asymptotic slopes at
small and large $R$ that one would expect from dimensional analysis
({\it i.e.} $\Sigma_*\propto\rho_{*}r$).
The definite integral of equation (A4) can be evaluated
as follows (\cite{gr80}):
\begin{equation}
J=c^{-\sigma}\ B(\gamma-0.5,0.5)\ _2F_1(\sigma,\gamma-0.5;\gamma;-b^{-1}),
\end{equation}
where $\gamma=\sigma+0.5\kappa$ is one-half the asymptotic density slope
as $r\rightarrow\infty$, $B$ is the beta function, and
$_2F_1$ is the hypergeometric function (\cite{gr80}).
We have empirically found that we can introduce a parameter
$\tau$, recast equation (A3) as
\begin{equation}
\Sigma_*={\rho_{*o}}r_c J'b^{-({{0.5\kappa}}-\tau)}
(1+b)^{-(\sigma+\tau-0.5)},
\end{equation}
where
\begin{equation}
J'=c^{-\sigma-\tau+0.5}\ B(\gamma-0.5,0.5)\ _2F_1(\sigma,\gamma-0.5;
\gamma;-b^{-1}),
\end{equation}
and determine a value of $\tau$ (for any $\kappa\le 1$) that results in a very
nearly constant $J'$
as a function of $R$ (with no
significant structure near $R=r_c$). Equation (A6) has the expected
slope as $R\rightarrow\infty$; and, we find
$\tau\rightarrow 0$ as $\kappa\rightarrow 0$ (the ``$\beta$ model''), and
$\tau\rightarrow 0.5$ as $\kappa\rightarrow 1$. We conclude that the
projection of a density profile of the form expressed by
equation (A1) is well-described
by equation (2) (at least for $a=2$), thus perhaps
lending some physical justification to its use.

\clearpage


\clearpage

 
\figcaption{$\beta_{\rm spec}$ vs. log $L_V$ for the DW sample (points,
with vertical lines representing 90\% confidence limits), and
best-fit $\beta_{\rm spec}$ -- log $L_V$ (dashed curve)
and log$(\beta_{\rm spec})$ -- log $L_V$ (dot-dashed curve)
linear regressions. See Section 4.5 for details.}

\figcaption{Comparison of luminosity density profiles, $l(r)$,
for ``deV'' (solid curve), ``HST'' (dotted curve), and ``Hernquist''
(dashed curve) stellar models
(see text). The radii are normalized to $r_{\rm max}=6r_e$, $l(r)$ to
the average luminosity density within $r_{\rm max}$, 
$l_{avg}=3L/4\pi{r_{\rm max}}^3$ where $L$ is the total stellar luminosity.}

\figcaption{$\beta_{\rm spec}$ {\it vs} velocity dispersion anisotropy
scale length, $s$, in units of the break radius, $r_{\rm br}$, for
HST ($\alpha=0$: solid curve, $\alpha=3$: dashed curve),
and Hernquist ($\alpha=0$: dotted curve, $\alpha=3$: dot-dashed curve)
stellar models. ($\alpha\equiv [M_{\rm dm}/M_*]_{r_{\rm max}}$.)}

\figcaption{(a) $\beta_{\rm spec}$ {\it vs} $r_{\rm dm}/r_{\rm br}$ for
$[M_{\rm dm}=3M_*]_{r_{\rm max}}$ (solid curve: HST stellar model,
dotted curve: Hernquist model).
(b) Baryon fractions at $r_{\rm br}$ (dashed curve for HST,
dot-dashed curve for Hernquist model)
and $r_e$  (solid curve for HST,
dotted curve for Hernquist model)
{\it vs} $r_{\rm dm}/r_{\rm br}$ for $[M_{\rm dm}=3M_*]_{r_{\rm max}}$.}

\figcaption{$\beta_{\rm spec}$ {\it vs} $[M_{\rm dm}/M_*]_{r_{\rm max}}$
for HST (solid curves), and Hernquist (dotted curves) stellar models
and $r_{\rm dm}=$0.5, 1.0, and $2.0r_e$
(upper, middle, and lower curves, respectively).}

\figcaption{(a) Minimum values of $r_{\rm dm}/r_{\rm br}$ for
HST and Hernquist models.
(b) Maximum values of baryon fractions within
(from top to bottom) $r_{\rm br}$, $r_e$, and $r_{\rm max}$ for
HST and Hernquist models.
The HST (Hernquist) model is represented by solid (dotted) curves.}

\figcaption{(a) Ratio of dark to luminous matter
within $r_{\rm max}$, {\it vs} dimensionless luminosity,
$\lambda\equiv L_V/L_0$ 
($L_0=5.2\times 10^{10}{h_{80}}^{-2}$L$_{V_\odot}$), for HST stellar models.
The dot-dashed curve denotes constant (within the virial radius)
baryon fraction ($f_{\rm baryon}=0.06$) model and CDM scaling of
dark matter concentration; dotted curve has $f_{\rm baryon}$ increasing
with optical luminosity; dashed curve has a steeper-than-CDM
scaling of concentration with dark halo mass. (b) Predicted 
variation of $\beta_{\rm spec}$ with dimensionless
luminosity. Line-types are as in (a). Data points and best-fit correlation
(solid curve) from Figure 1 are re-plotted. (For clarity, three
galaxies where $\beta_{\rm spec}$ is uncertain to $>50$\% are omitted 
from this and the following figure.)}

\figcaption{(a) $\beta_{\rm spec}$ calculated using pairs ($L_V$, $\sigma$)
as measured for the DW sample 
and CDM scaling of
dark matter concentration with constant $f_{\rm baryon}$
vs. observed
$\beta_{\rm spec}$. (b) Same as (a) for model where
$f_{\rm baryon}$ increases
with optical luminosity. (c) Same as (a) for model with
steeper-than-CDM
scaling of concentration with dark halo mass. The solid line
represents $\beta_{\rm spec}=\beta_{\rm obs}$.}

\figcaption{(a) Mass {\it vs} $L_V$ (in solar units) within,
from bottom to top,
$r_e$, $r_{\rm max}=6r_e$ (stars only), $r_{\rm max}$ (total), and $r_{200}$.
The solid curves show the
mass at $r_e$ and $r_{\rm max}$ from weak gravitational lensing studies.
(b) Same as (a) for mass-to-light ratio (in solar units) at
(from lowest
to uppermost curves): $r=0$, $r_e$, $r_{\rm max}$, and $r_{200}$. (c) Same as (a)
for baryon fraction {\it vs} $L_V$ at (from uppermost
to lowest curves) $r=r_{\rm br}$, $r_e$, $r_{\rm max}$, and $r_{200}$.
Dotted and dashed line-types have the same connotations
as in Fig. 7.}

\figcaption{1-d velocity dispersion distributions, assuming isotropic orbits,
for an $L_V=L_0=5.2\times 10^{10}{h_{80}}^{-2}$L$_{V_\odot}$ galaxy.
``Mass loss'' and ``non-CDM'' models are denoted 
by dotted and solid curves, respectively, for the stellar profiles and
dot-dashed and dashed curves, respectively, for the dark matter profiles.
$[M_{\rm dm}/M_*]_{r_{\rm max}}=1.5(1.3)$ and $r_{\rm dm}/r_e=4.1(12)$
for the mass loss (non-CDM) models at this luminosity, and the
(observational and model) value of $\beta_{\rm spec}$ is 0.62.}

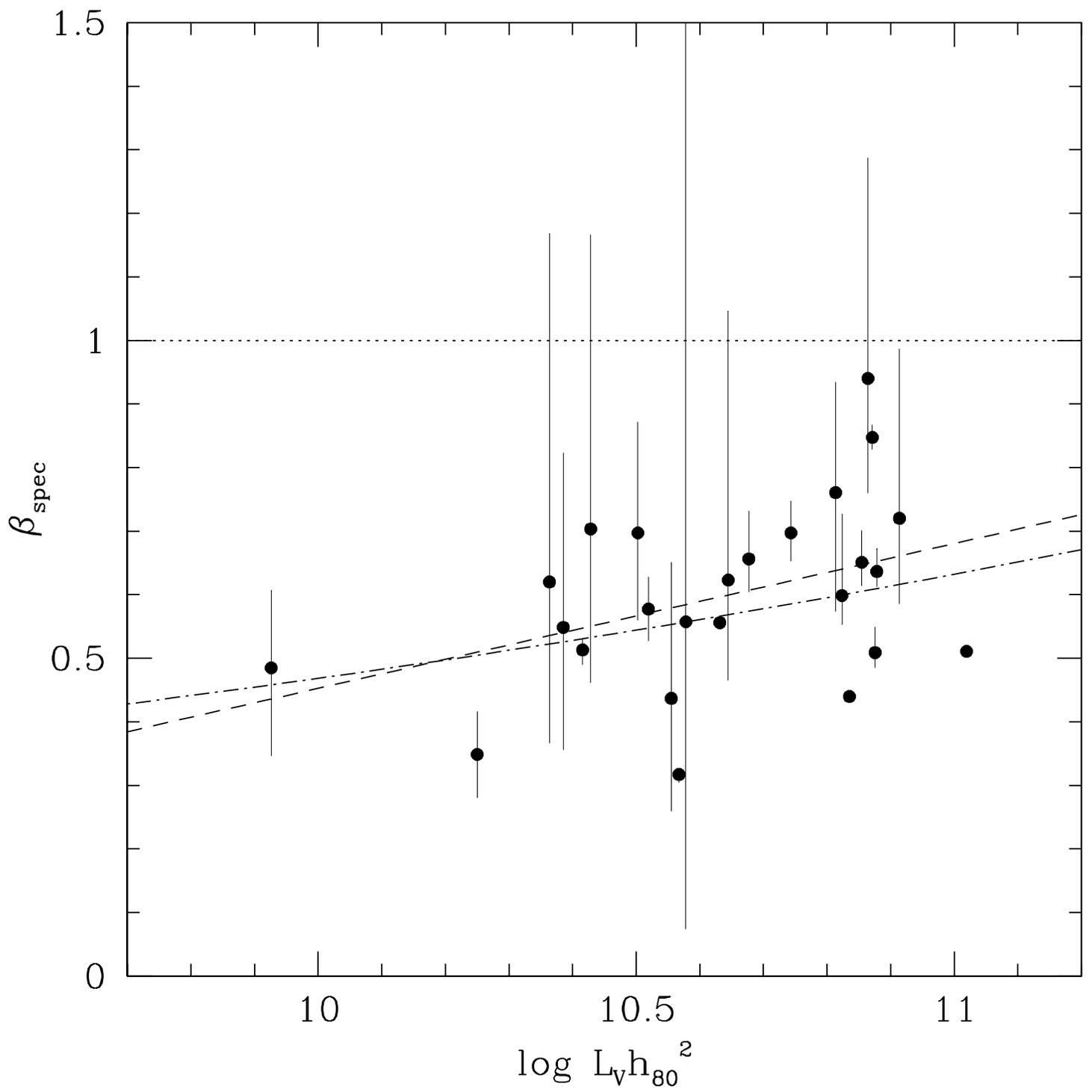
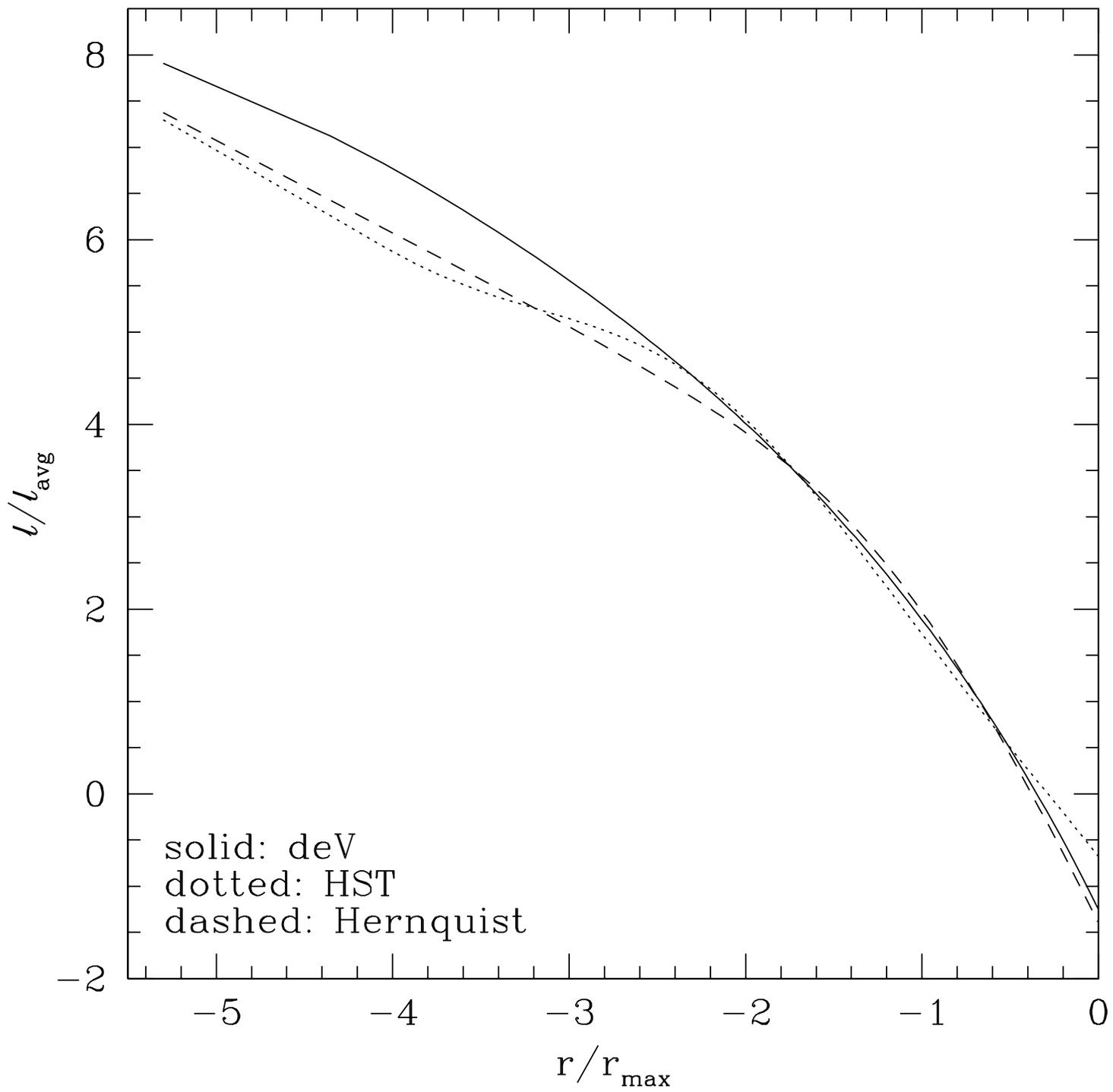
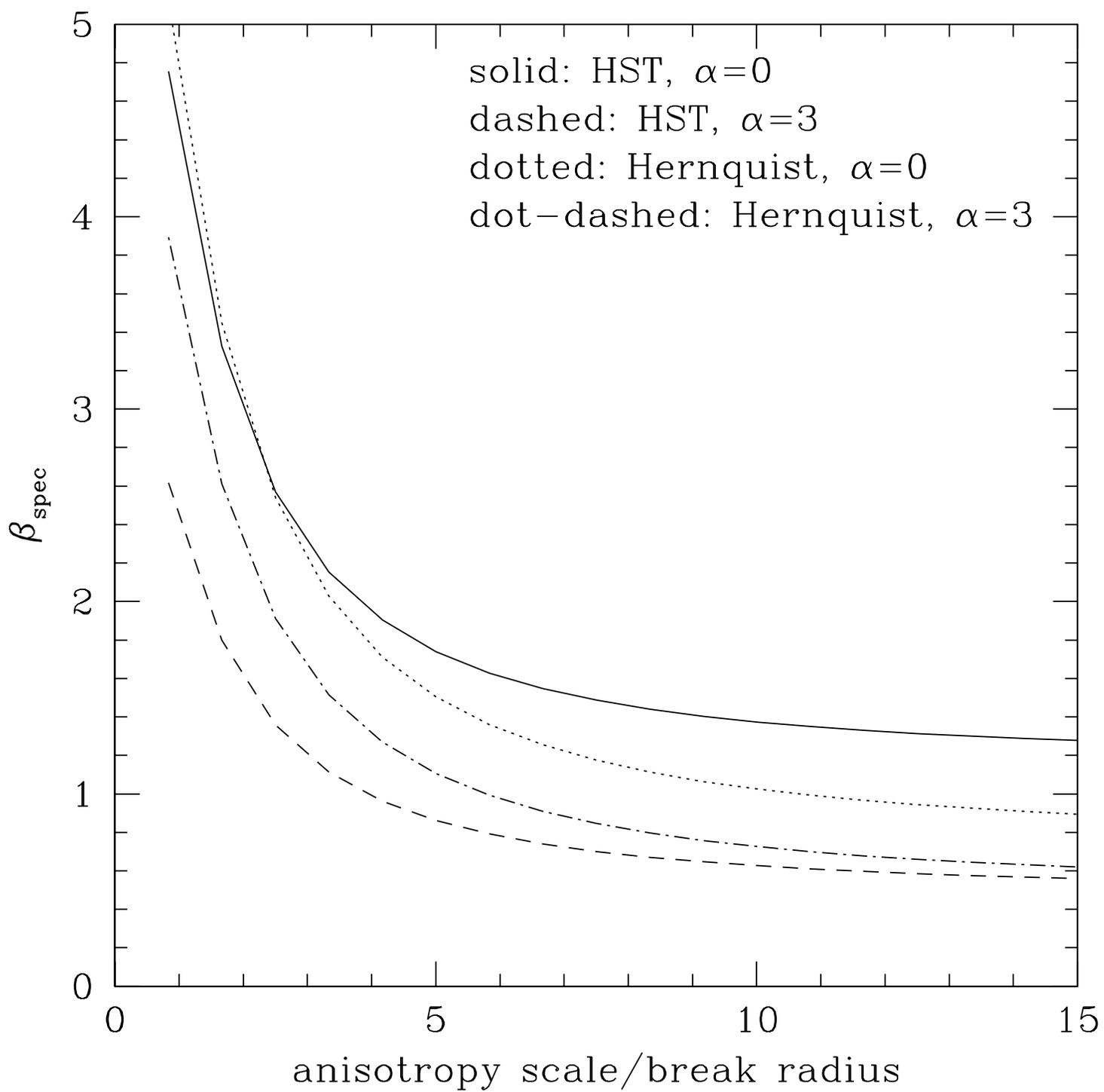
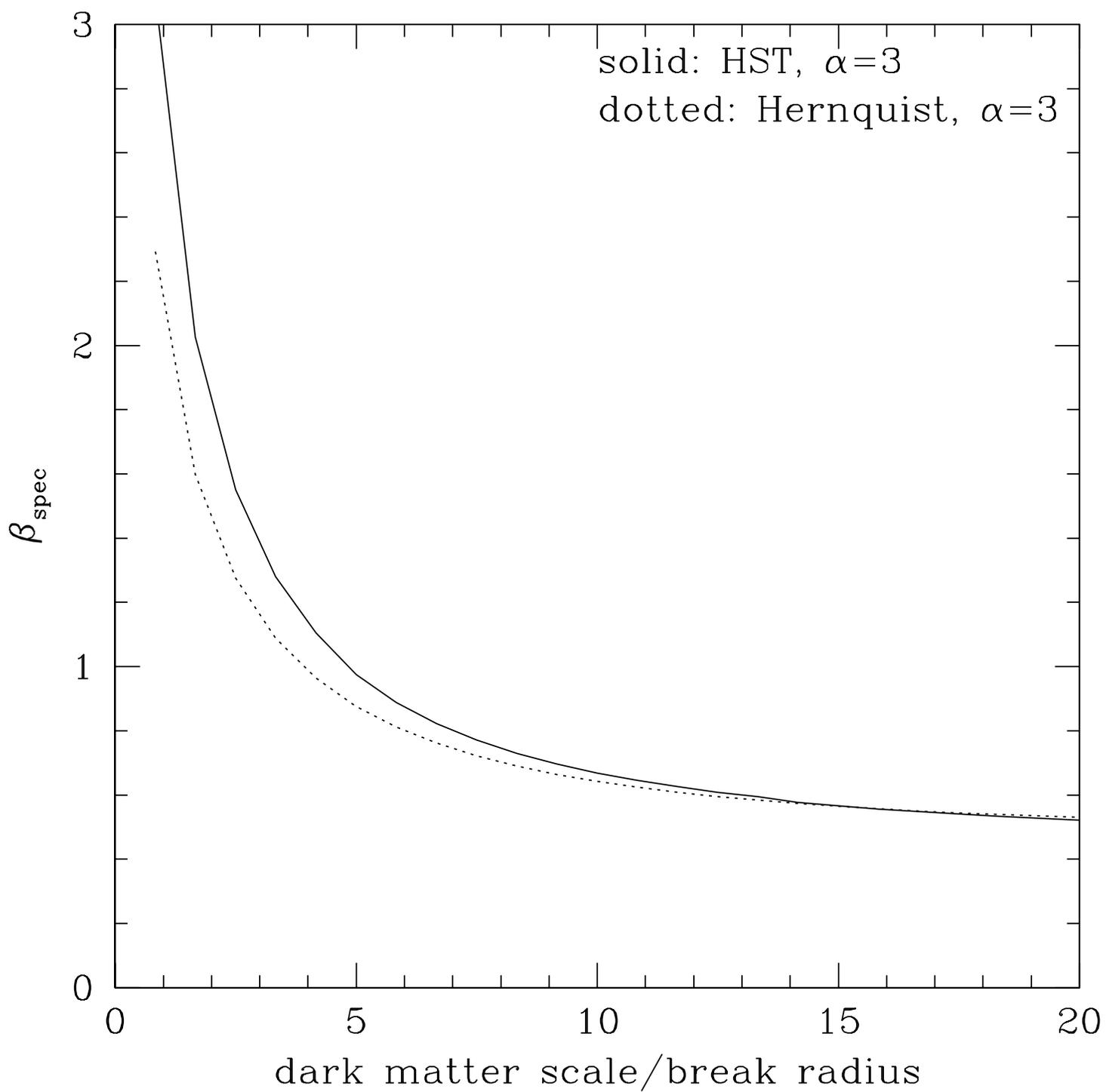
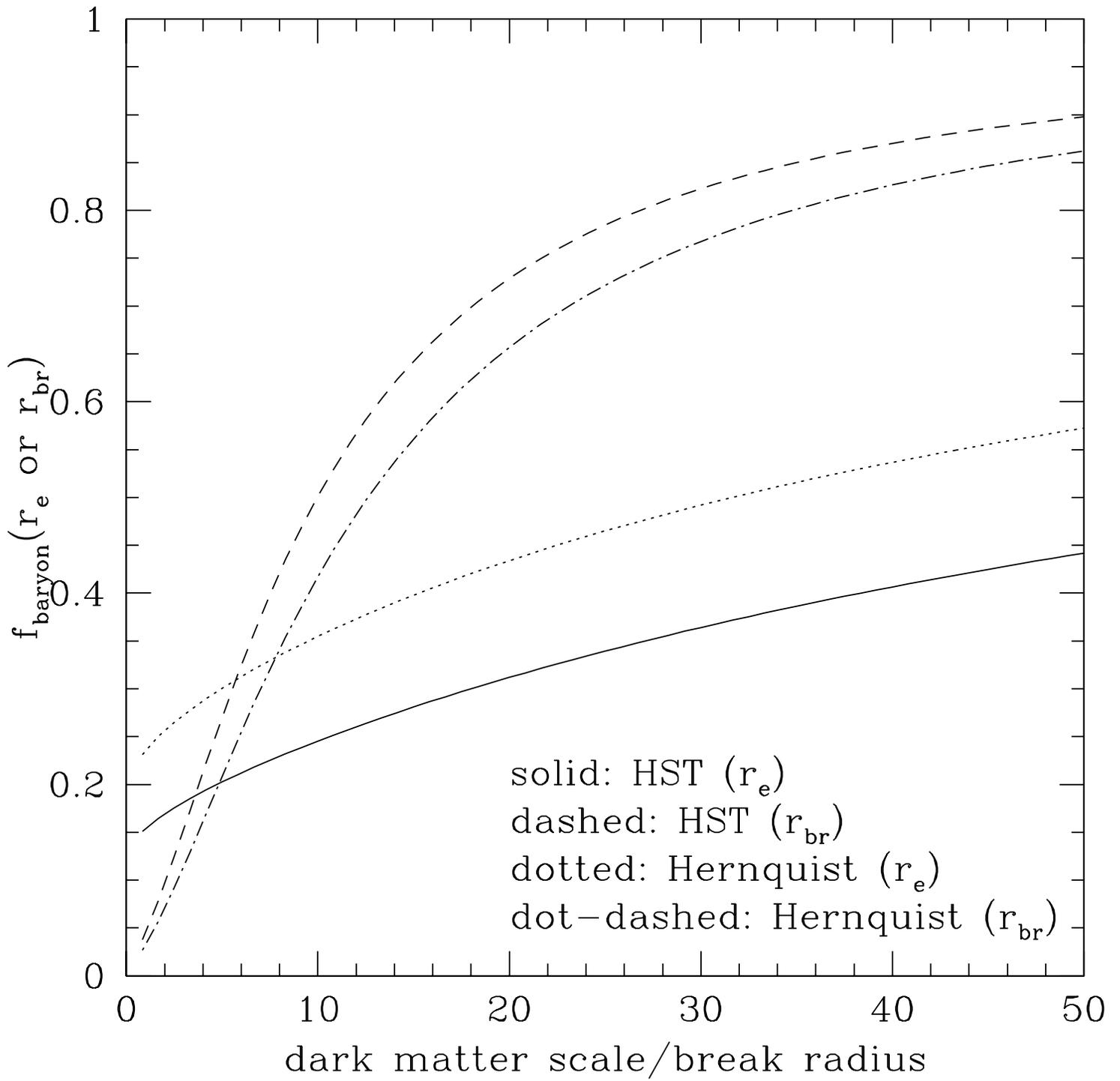
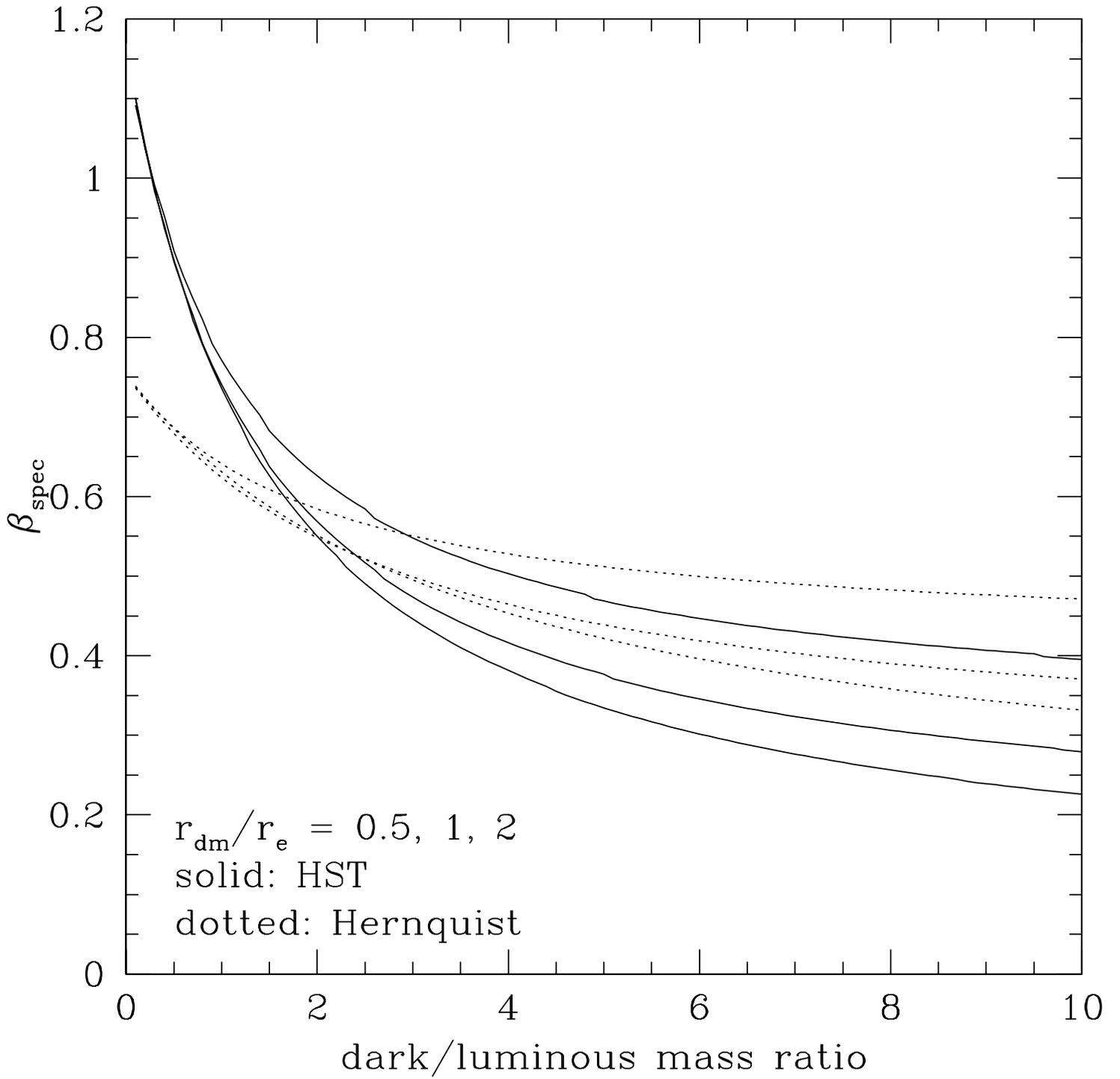
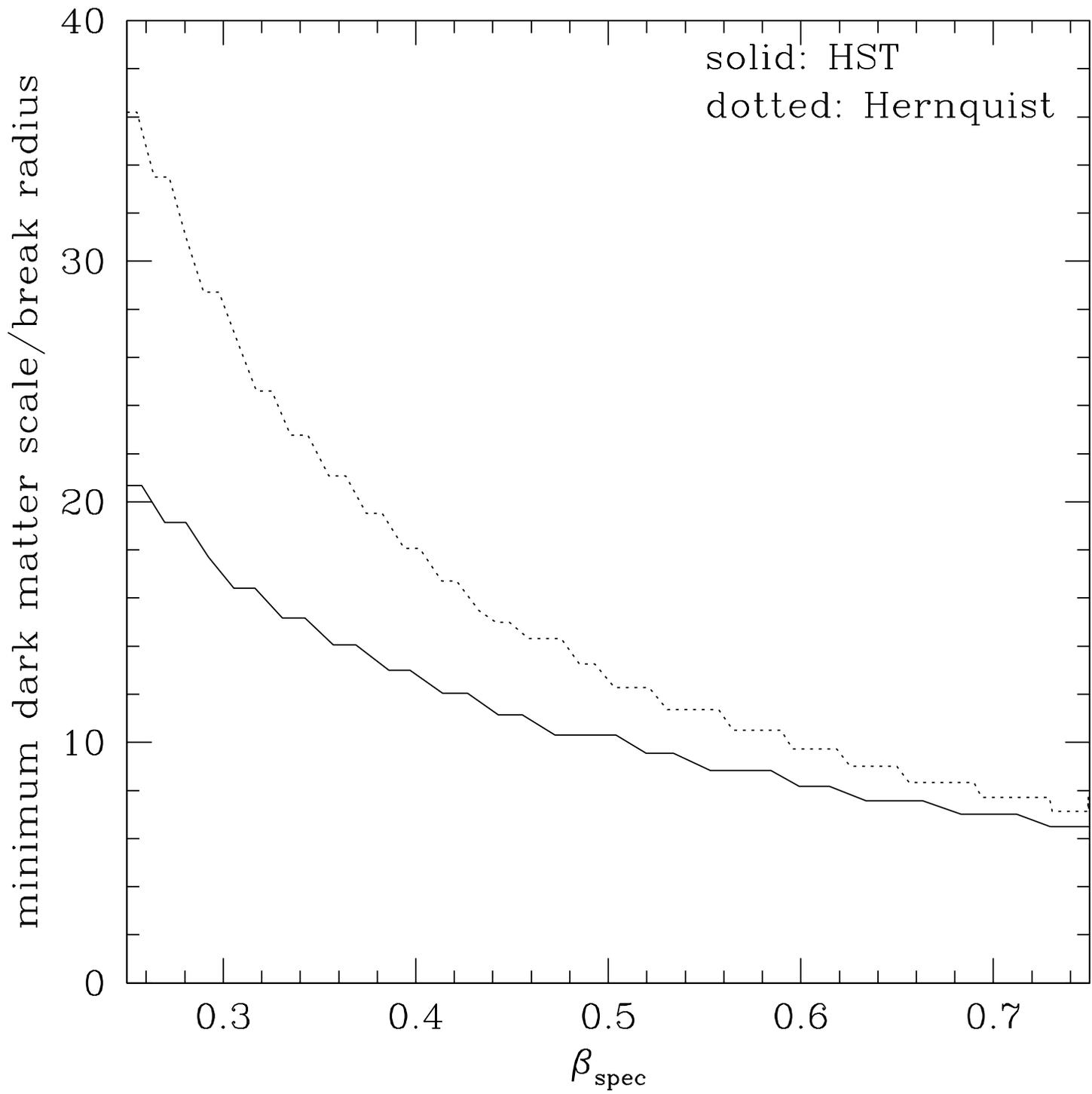
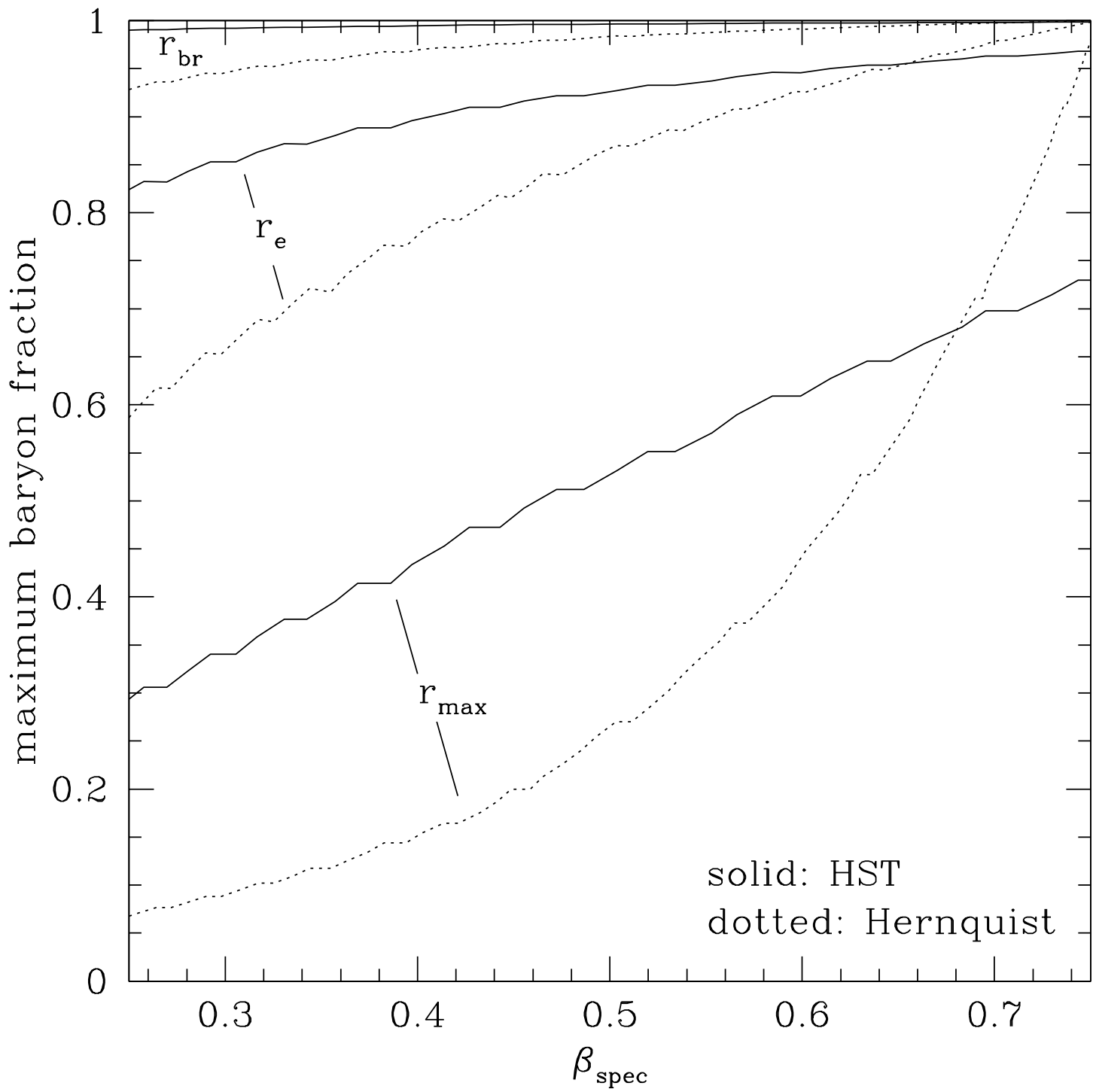
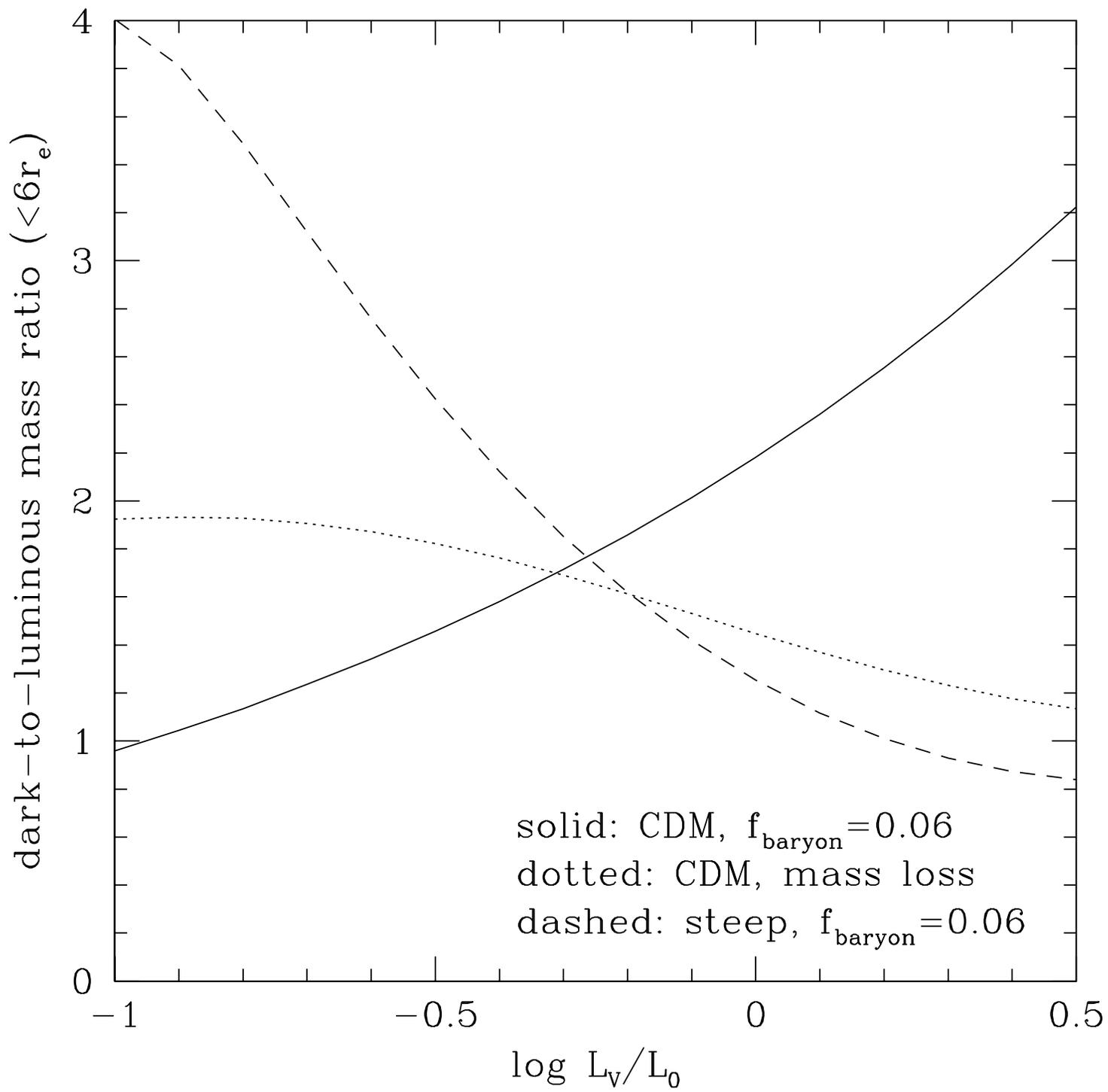
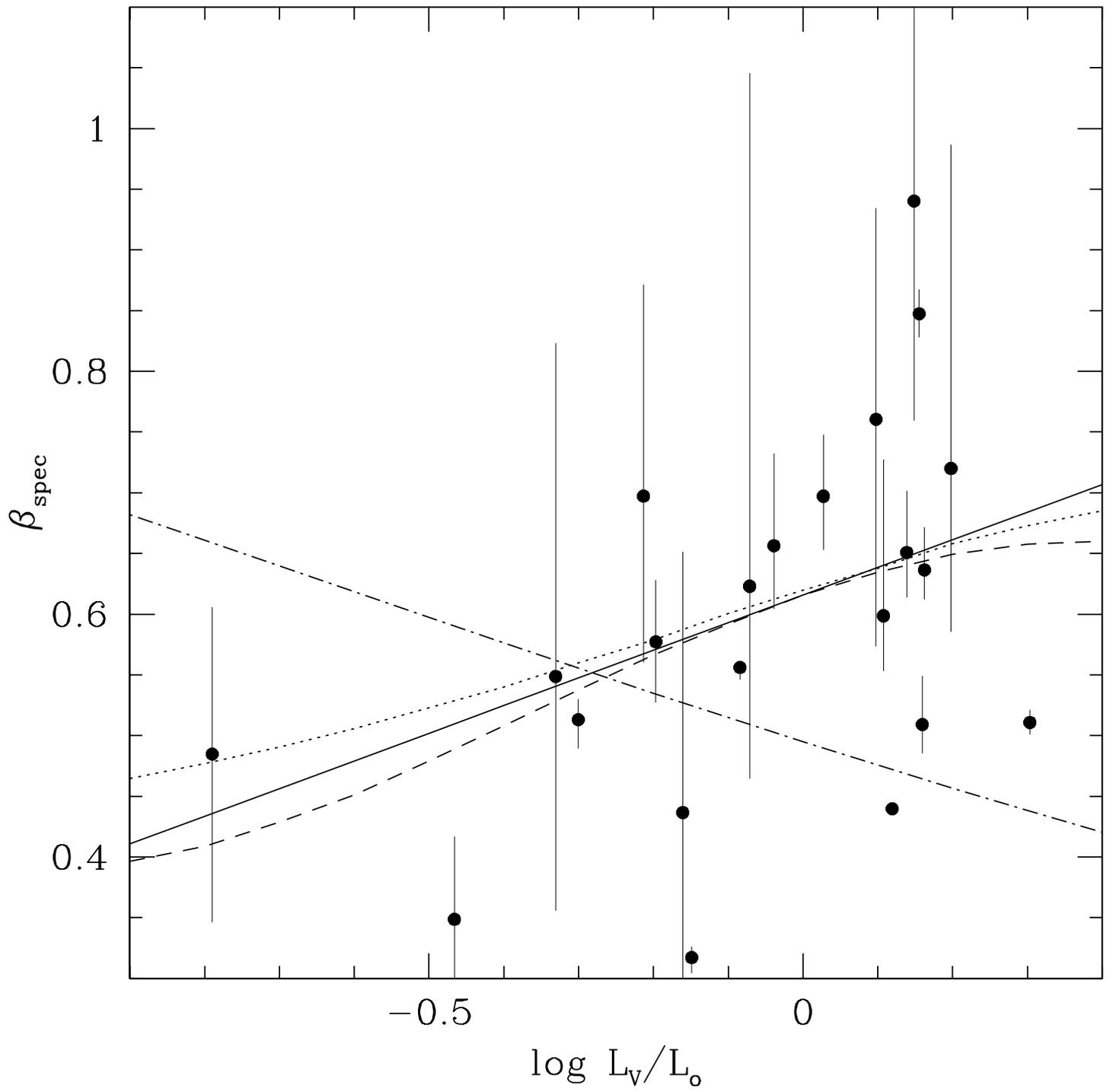
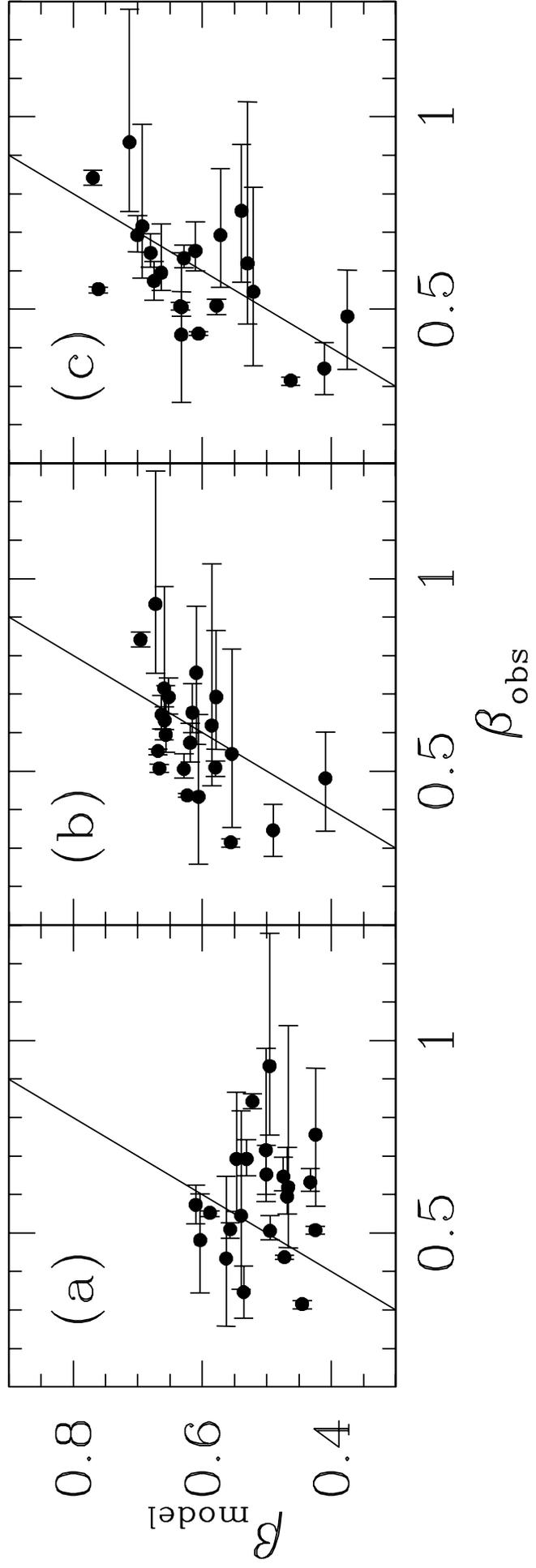
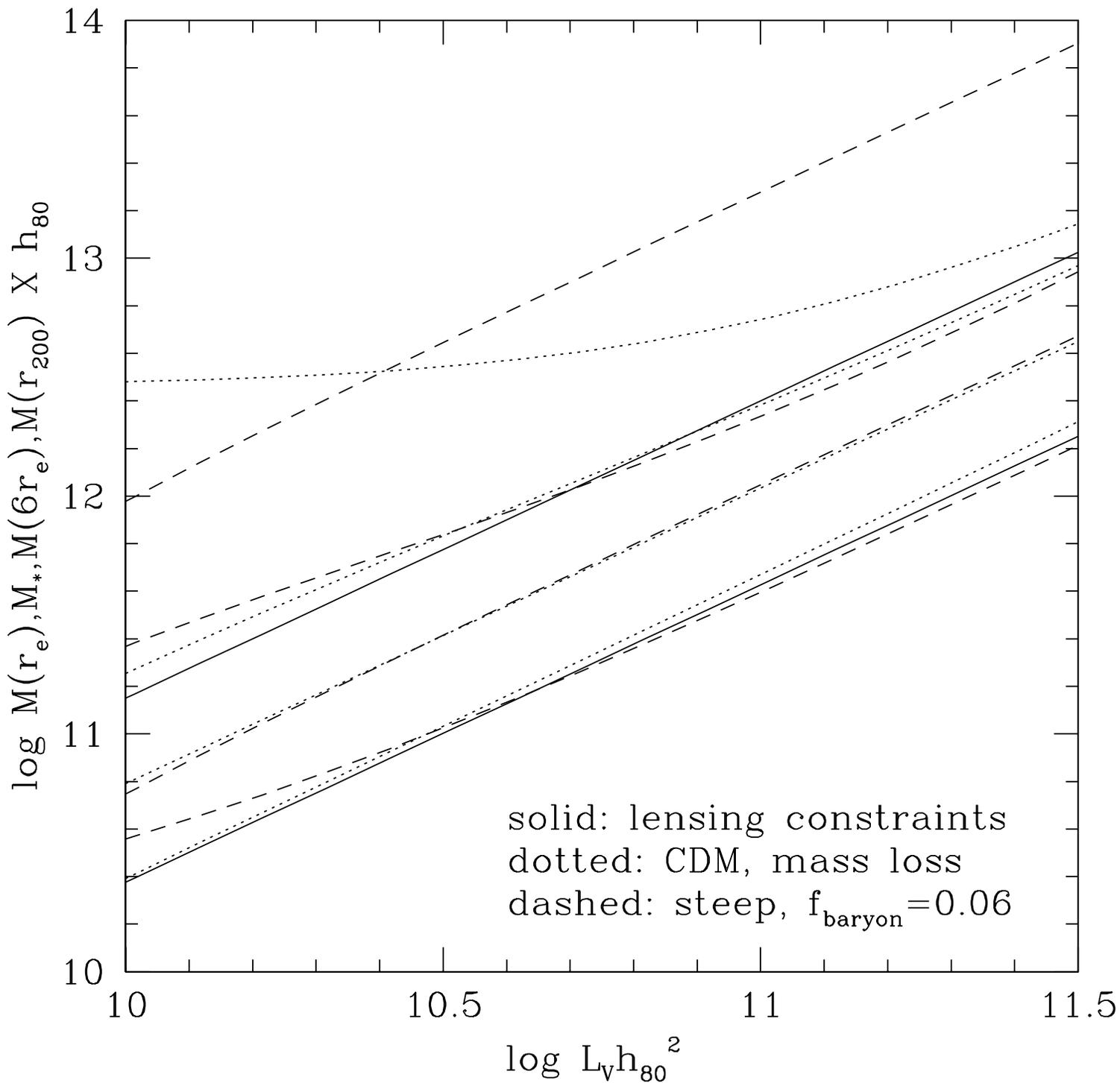
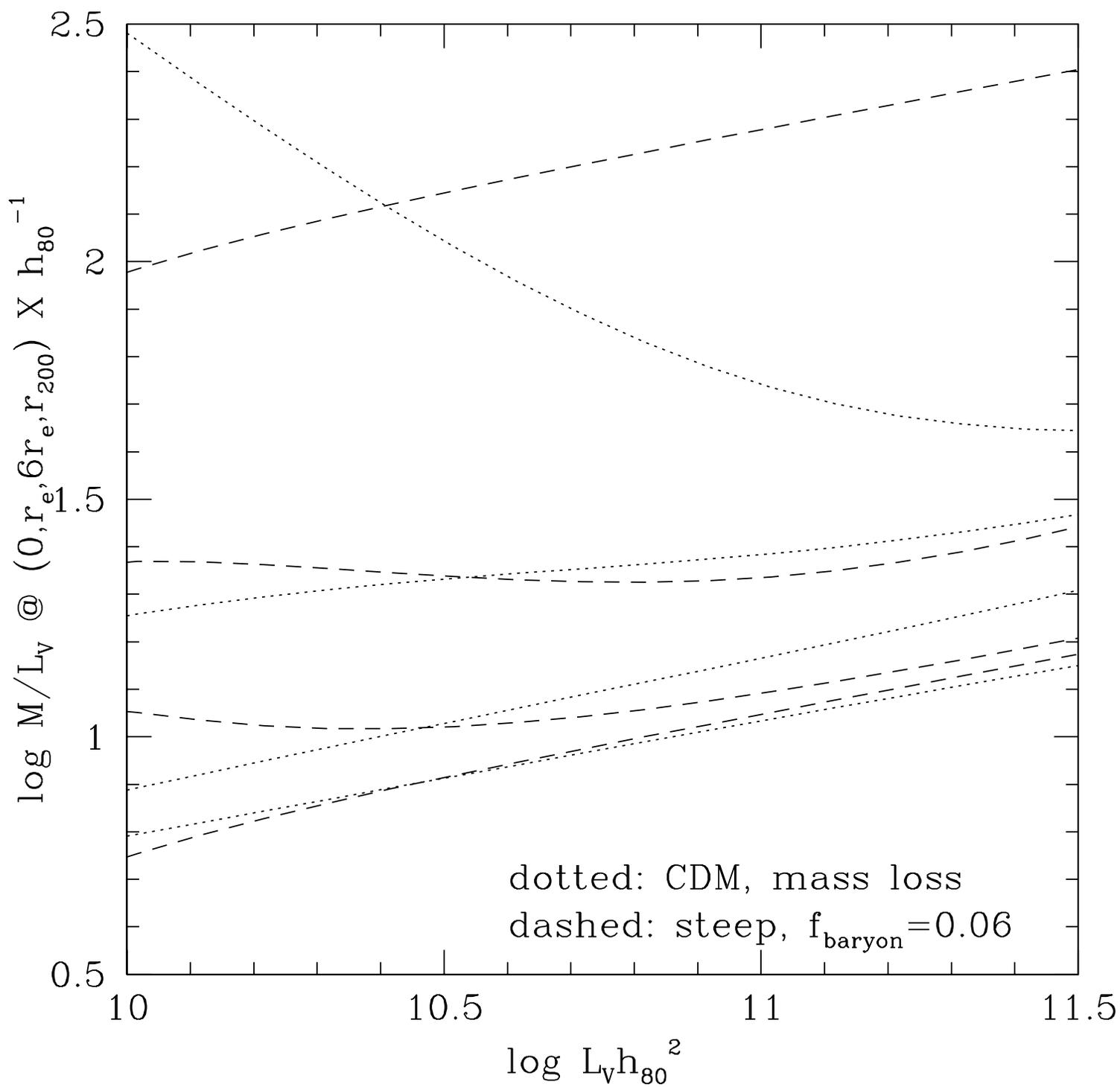
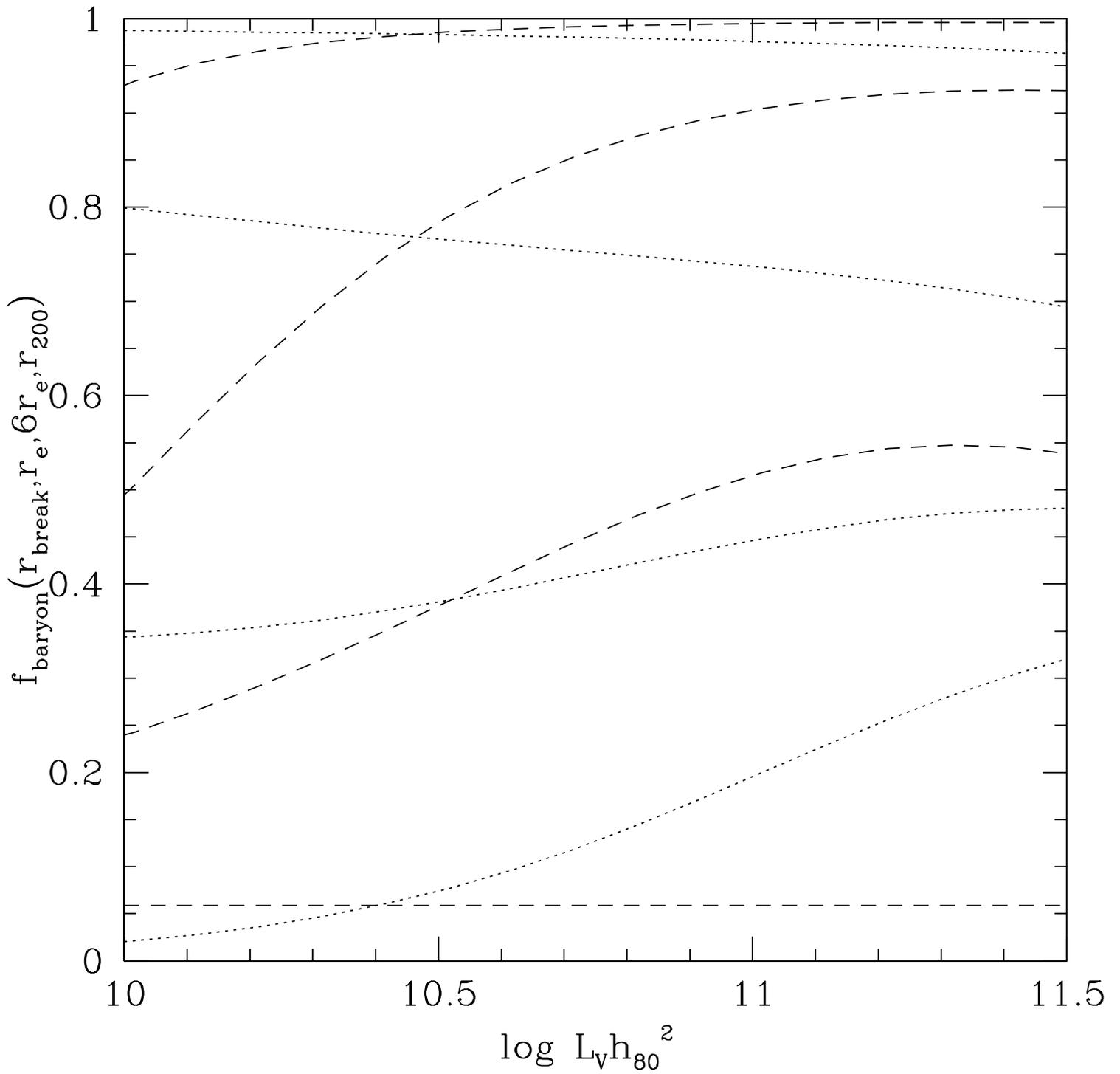
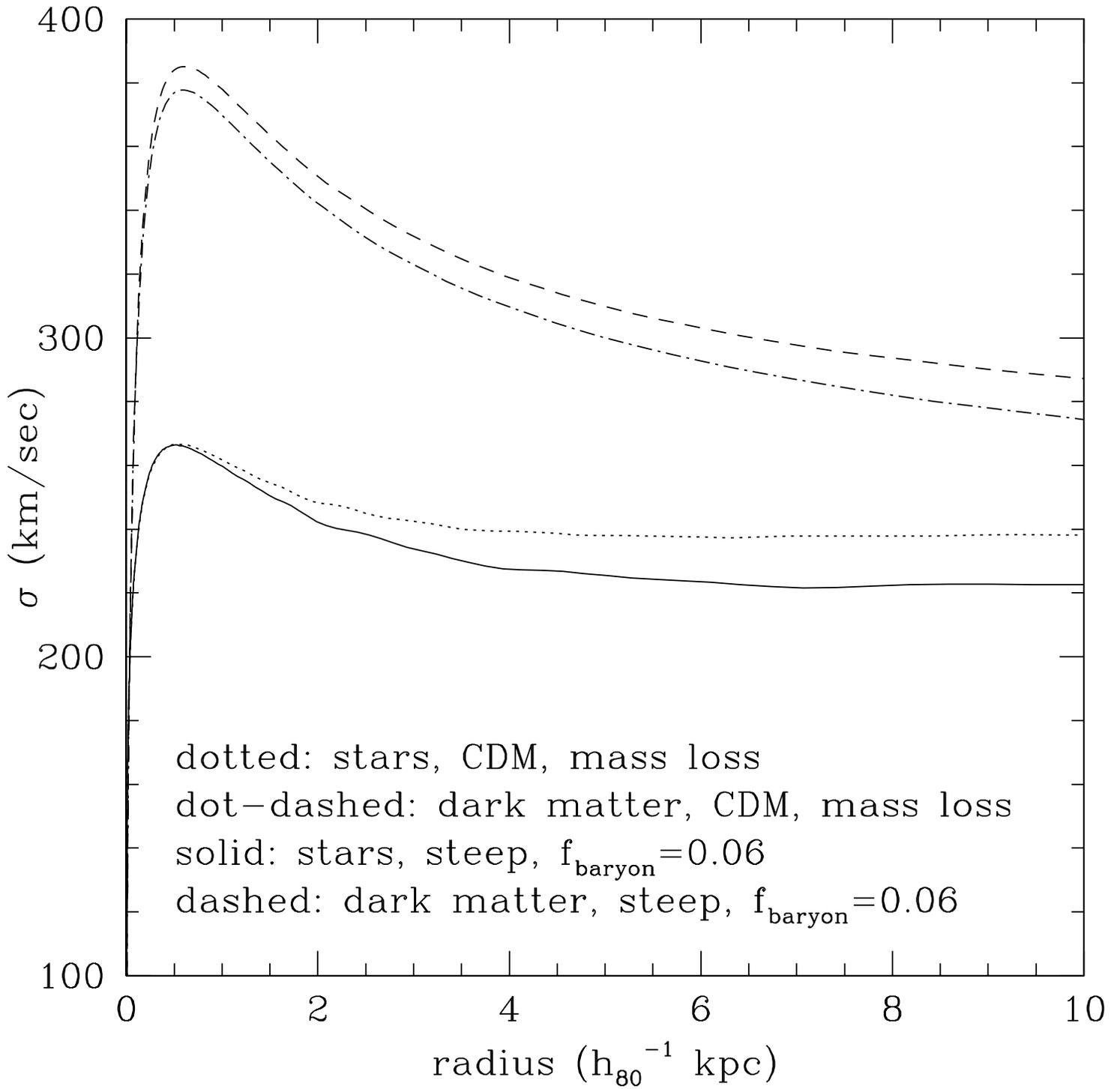
\end{document}